\documentclass[12pt,aps,showpacs,preprintnumbers,amsmath,amssymb,%
prd,a4paper,floatfix,nofootinbib,superscriptaddress,showkeys]{revtex4-1} 
\usepackage{graphicx} % Used for includegraphics
\usepackage{array}

\graphicspath{{./Figures/}} % location for color fig.

\usepackage{bbm}

\newcommand{\be}{\begin{equation}}

\newcommand{\ee}{\end{equation}}

\newcommand{\bea}{\begin{eqnarray}}

\newcommand{\eea}{\end{eqnarray}}

\newcommand{\Tr}{\mbox{Tr}}

\begin{document}

\title{Topological order and the vacuum of Yang-Mills theories.}

\author{G.~Burgio}
\author{H.~Reinhardt}
\affiliation{Institut f\"ur Theoretische Physik, Auf der Morgenstelle 14, 
72076 T\"ubingen, Germany}

\begin{abstract}
%--------------
We study, for $SU(2)$ Yang-Mills theories discretized on a lattice, a 
non-local topological order parameter, the center flux ${{z}}$. We show
that: i) well defined 
topological sectors classified by $\pi_1(SO(3))=\mathbb{Z}_2$ can only exist 
in the ordered phase of ${{z}}$;
ii) depending on the dimension $2 \leq d\leq 4$
and action chosen, the center flux exhibits a critical 
behaviour sharing striking features with the 
Kosterlitz-Thouless type of transitions, although 
belonging to a novel universality class; iii) such critical behaviour 
does not depend on the temperature $T$. Yang-Mills theories can thus 
exist in two different continuum phases, characterized by an either 
topologically ordered or disordered vacuum; this reminds of a quantum phase
transition, albeit controlled by the choice of symmetries and not by a 
physical parameter. 
\end{abstract}

\maketitle

\section{Introduction}
\label{sec:intro}

Of all ideas applied to the confinement problem in non-abelian Yang-Mills 
theories \cite{Jaffe:2000ne,Greensite:2011zz} the most popular still involve 
topological degrees of freedom of some sort 
\cite{Wu:1967vp,Nielsen:1973cs,'tHooft:1974qc,%
Mandelstam:1974vf,Polyakov:1976fu,Goddard:1976qe,Mandelstam:1978ed,%
Weinberg:1979zt,Seiberg:1994rs}. Among these center vortices 
\cite{'tHooft:1977hy,'tHooft:1979uj,Mack:1981gy} have enjoyed broad attention, 
in 
particular in the lattice literature. Although most of the effort 
was put in dealing with gauge fixed schemes,\footnote{See e.g. 
Refs.~\cite{DelDebbio:1996mh,Langfeld:1997jx} for early results and 
Ref.~\cite{Greensite:2011zz}, Chapts.~6,~7 for a comprehensive review. The 
goal here is to isolate some relevant degrees of freedom, usually called
P-vortices, assumed to be related to 't Hooft's topological excitations; we 
will comment on this in Sec.~\ref{sec:con}.} some investigations 
actually attempted to tackle the problem in a gauge invariant way
\cite{Kovacs:2000sy,Hart:2000en,deForcrand:2000fi,deForcrand:2001nd,%
Burgio:2006dc,Burgio:2006xj,vonSmekal:2012vx} and are therefore
directly related to 't Hooft's original idea.

Non-abelian gauge fields transform under the group's adjoint representation, 
$SU(N)/\mathbb{Z}_N$.
Such group is not simply connected, with a non trivial first homotopy class:
\be
\pi_1(\frac{SU(N)}{\mathbb{Z}_N}) = \mathbb{Z}_N\,,
\label{eq:homotopy}
\ee
the center of $SU(N)$. 
Following Ref.~\cite{'tHooft:1979uj}, let us consider the Euclidean 
Yang-Mills theory on a $d$-dimensional torus,\footnote{See e.g. 
Ref.~\cite{GonzalezArroyo:1997uj} for an extensive introduction to the 
subject.} i.e. with all directions compactified, 
and choose one of the $d$ Euclidean directions as time. If large
gauge transformations classified by Eq.~(\ref{eq:homotopy}) induce a 
super-selection rule, we can decompose the physical Hilbert space ${\cal{H}}$ 
of gauge invariant states \cite{Burgio:1999tg} in sub-spaces 
${\cal{H}}_{\vec{k},\vec{m}}$ labeled by 
topological indices, the $\mathbb{Z}_N$ electric and magnetic fluxes (vortices) 
$\vec{k} = (k_1,\ldots,k_{n_t})$ and $\vec{m} = (m_1\ldots,m_{n_s})$ 
\cite{'tHooft:1979uj}:
\be
{\cal{H}} = \bigoplus_{{k}_i,{m}_j=0}^{N-1} {\cal{H}}_{\vec{k},\vec{m}}\,.
\label{eq:hilbert}
\ee
Here $n_t = d-1$ counts the space-time and $n_s = \frac{(d-1)(d-2)}{2}$ the 
space-space planes and $k_i, m_i \in \mathbb{Z}_N$. As
't Hooft pointed out, a sufficient condition for 
confinement is realized if the low-temperature phase of pure Yang-Mills
theories corresponds to a superposition of all (electric) sectors, while above 
the deconfinement transition such $\mathbb{Z}_N$ symmetry must get broken to 
the trivial one \cite{'tHooft:1977hy,'tHooft:1979uj}.\footnote{Magnetic 
sectors, on the other hand, can remain unbroken and be responsible for 
screening effects.}

One can check such scenario by
calculating, e.g. in lattice simulations, how the free energy for 
$\mathbb{Z}_N$ flux creation:
\be
F(\vec{k}) = \Delta\, {\cal{U}}_{\vec{k}} -T\, \Delta\, {\cal{S}}_{\vec{k}} = 
-\log{\frac{Z(\vec{k})}{Z(\vec{0})}}
\label{eq:free_en}
\ee 
changes with the temperature $T$ across the deconfinement transition. 
Here $ \Delta\, {\cal{U}}_{\vec{k}}$ is the energy (action) cost
to generate the $\vec{k}^{\rm{th}}$ (electric) vortex from the vacuum, 
$\Delta\, {\cal{S}}_{\vec{k}}$ the corresponding
entropy change and $Z(\vec{k})$ the partition function restricted to the 
topological sector labeled by $\vec{k}$.\footnote{In the deconfined phase all 
electric sectors must be suppressed relatively to the trivial one,
while in the confined phase all $\vec{k}$ and $\vec{m}$ should be equally 
probable.} For the one-vortex sector, 
$F$ is nothing but the free energy of a maximal 't Hooft loop,\footnote{For
a representation of the 't Hooft loop in the continuum see 
Ref.~\cite{Reinhardt:2002mb}} giving a 
confinement criterion dual to Wilson's: in the thermodynamic 
limit $F$ should vanish in the confined phase while it should diverge as 
$\tilde{\sigma}(T)\, L^2$ above the deconfinement temperature $T_c$, where 
$\tilde{\sigma}(T)$ is the dual string tension
\cite{'tHooft:1979uj,Greensite:2011zz,Kovacs:2000sy,Hart:2000en,%
deForcrand:2000fi,deForcrand:2001nd,deForcrand:2002vs,Burgio:2006dc}. In other
words, a perimeter law for the Wilson loop implies an area law for the 
't Hooft loop and vice-versa \cite{'tHooft:1977hy,'tHooft:1979uj}.

Of course, when considering the theory at $T=0$, the distinction
between electric and magnetic fluxes is artificial. In this case 
{\it all} the $N^{\frac{d (d-1)}{2}}$ topological sectors must be taken into account 
when establishing whether $\mathbb{Z}_N$-symmetry is unbroken, i.e. whether the 
vacuum $|\Psi_0\rangle$ is indeed a symmetric superposition of 
states belonging to ${\cal{H}}_{\vec{k},\vec{m}}$:\footnote{Actually, in virtue of 
cubic symmetry, one can regard the sub-spaces 
${\cal{H}}_{\vec{k},\vec{m}}$ with indices equal up to a permutation as 
equivalent and recombine them in Eq.~(\ref{eq:vac}) into weights given 
by their combinatorial multiplicity 
\cite{Kovacs:2000sy,Hart:2000en,deForcrand:2000fi}.}
\be
|\Psi_0\rangle = \sum_{\vec{k},\vec{m}} |\Psi_0^{\vec{k},\vec{m}}\rangle\,.
\label{eq:vac}
\ee 
In the following we will use either definition, depending on whether we are 
considering the $T=0$ or the $T>0$ case.

The above ideas generalize naturally to the lattice discretization 
of Yang-Mills theories; the specific action used plays however a key
role in their actual implementation. If one wishes to preserve the 
symmetries of the continuum theory the natural choice should fall on 
a discretization transforming under the ``correct'' group $SU(N)/\mathbb{Z}_N$. 
One possibility among many (see e.g. Ref.~\cite{Halliday:1981tm}) is given by 
the adjoint Wilson action with periodic boundary conditions 
\cite{deForcrand:2002vs}:
\begin{equation}
S_A = \beta_{A}\sum_{P}\left(1-\frac{1}{N^2-1}\Tr_{A}\,U_{P}\right)\,, 
\label{eq:adj_action}
\end{equation}
where $U_P$ is the standard plaquette. For $N=2$ it was indeed shown in 
Refs.~\cite{Burgio:2006dc,Burgio:2006xj,Barresi:2006gq} that 
for simulations based on Eq.~(\ref{eq:adj_action}): 

\noindent i) $\mathbb{Z}_2$ topological sectors are well 
defined in the continuum limit, both below and above $T_c$, i.e. 
the decomposition in Eq.~(\ref{eq:hilbert}) holds; 

\noindent ii) the partition function 
$Z_A = \int \exp{(-S_A)}$ dynamically includes all sectors; 

\noindent iii) in the deconfined phase all non trivial sectors are suppressed,
while as $T\to 0$ all sectors are equivalent, i.e. the vacuum can be described 
by Eq.~(\ref{eq:vac}). 

\noindent The main difficulty of such setup lies of course in the 
implementation of an algorithm capable of tunneling ergodically among all 
vortex topologies. 
Simulations are therefore quite demanding: reaching enough statistics to
check whether the symmetry among sectors postulated in 
Eq.~(\ref{eq:hilbert}) remains unbroken from $T = 0$ all the way up to $T_c$ is 
difficult; the evidence given in Refs.~\cite{Burgio:2006dc,Burgio:2006xj} 
seems to point to a more complicated picture. 

Alternatively, universality \cite{Svetitsky:1982gs} 
should allow the use of the fundamental Wilson action: 
\begin{equation}
S_F = \beta_{F}\sum_{P} \left(1-\frac{1}{N}\Re\left[\Tr_{F}\,U_{P}\right]\right)\,,
\label{eq:fund_action}
\end{equation}
which is the quenched (mass $\to \infty$) limit of the physical action
coupling Yang-Mills theories to fundamental fermions, e.g. full QCD.
In this case, however, some care must be taken in defining a 
$SU(N)/\mathbb{Z}_N$ invariant theory.
Indeed, in the presence of fundamental fermions 
the topological classification of Eq.~(\ref{eq:homotopy}) breaks 
down.\footnote{See e.g. Ref.~\cite{Cohen:2014swa} for a recent 
discussion.}  
The extension to full QCD has been indeed one of the main obstacles 
in establishing
the 't Hooft vortex picture as a viable model for confinement. We will comment
on this in Sec.~\ref{sec:con}; for the moment, let us note that 
one can still introduce vortex topological sectors ``statically'' by simply 
imposing twisted boundary conditions
\cite{deForcrand:2000fi,deForcrand:2001nd,deForcrand:2002vs}.\footnote{Such 
topological boundary conditions, 
relevant e.g. in investigations of large $N$ reduction 
\cite{GonzalezArroyo:1997uj,GonzalezArroyo:1982hz,Perez:2014sqa}, allow 
adjoint fermions but no fundamental ones. Flavour
twisted boundary conditions, on the other hand, are well established in full
QCD \cite{Sachrajda:2004mi}.}  
We should then be able to reconstruct the ``full'' partition function ${Z_F}$ 
by taking the weighted sum of all partition functions 
$Z_F(\vec{k},\vec{m}) = \int \exp{(-S_F({\vec{k},\vec{m}}))}$ with 
boundary conditions 
corresponding to the sector labeled by $\vec{k}$ and $\vec{m}$ 
\cite{deForcrand:2000fi,deForcrand:2001nd}. Since each $Z_F(\vec{k},\vec{m})$ 
must be determined via independent simulations, their relative weights can 
only be calculated through indirect 
means.\footnote{See e.g. Ref.~\cite{vonSmekal:2012vx}, Chapt.~3 for
a detailed review of the methods involved.}
Still, such simulations are computationally more efficient than in the 
$Z_A$ case and have therefore been the method of choice in most investigations 
of Eq.~(\ref{eq:free_en}) 
\cite{Kovacs:2000sy,Hart:2000en,deForcrand:2000fi,deForcrand:2001nd,%
vonSmekal:2012vx}.

Investigations using Eq.~(\ref{eq:fund_action}) rely on the 
assumption that fixing the boundary conditions is enough
to ensure that the Hilbert-space decomposition defined in 
Eq.~(\ref{eq:hilbert}) works. However, 
it is well known that upon discretization of Yang-Mills theories
$\mathbb{Z}_N$ magnetic monopoles are generated at strong coupling
\cite{Lubkin:1963zz,Mack:1978rq,Mack:1979gb,Halliday:1981te,Halliday:1981tm,%
Coleman:1982cx}, causing bulk phenomena in the $\beta_F - \beta_A$ 
phase diagram. 
Now, since the $\mathbb{Z}_N$ fluxes defining our topological sectors live on 
the co-set of a two dimensional plane, they have a simple 
geometrical interpretation: they are described in $d=4$ by a {\it closed} 
world-sheet, i.e. they are string-like objects, and in $d=3$ by a {\it closed}
world-line, i.e. particle-like. On the other hand, topological lattice 
artifacts as the above mentioned $\mathbb{Z}_N$ monopoles are themselves 
sources of $\mathbb{Z}_N$ flux: in $d=4$ 
they will be particle-like objects, their {\it closed} world-lines bounding 
{\it open} $\mathbb{Z}_N$ flux world-sheets, while in $d=3$ they will be 
instanton-like objects and will be end-points of {\it open} $\mathbb{Z}_N$ 
flux lines 
\cite{Lubkin:1963zz,Mack:1978rq,Mack:1979gb,Coleman:1982cx,deForcrand:2002vs}.
$\mathbb{Z}_N$ monopoles are therefore in one to one correspondence with 
{\it open} center vortices; in other words, universality between the
fundamental and adjoint actions
can only be invoked when just {\it closed}, i.e. truly topological 
${\mathbb{Z}}_N$ vortices winding around the compactified directions can form.
Notice how in $d=2$, where no $\mathbb{Z}_N$ monopoles can exist, 
$\mathbb{Z}_N$ fluxes are instanton type objects. The distinction between open 
and closed vortices is in this case blurred, but in a non-ergodic setup 
it can eventually be made through the flux allowed by the boundary 
conditions chosen.

The above discussion has a straightforward consequence. If one could 
``measure'' 
whether open ${\mathbb{Z}}_N$ vortices are absent in a given discretization, 
i.e. whether only topological vortices can be generated from the vacuum, 
there would be no need to monitor $\mathbb{Z}_N$ monopoles to
establish universality between $S_A$ and $S_F$ in the 
first place, since these must be absent 
anyway. This would have two advantages: first, such criterion 
could be generalized to $d=2$. Second, 
absence of lattice artifacts, whether for $S_F$, for $S_A$ or for both, 
would get ``promoted'' to a necessary condition for the super-selection rule 
of Eq.~(\ref{eq:hilbert}), and hence for the conjectured vacuum symmetry  
of Eq.~(\ref{eq:vac}), to be realized. Indeed, 
consider states belonging to distinct topological sectors labeled by the 
indices $k$, $k' \in \mathbb{Z}_N$.
\begin{figure}
\begin{center}
\includegraphics[width=0.7\textwidth]{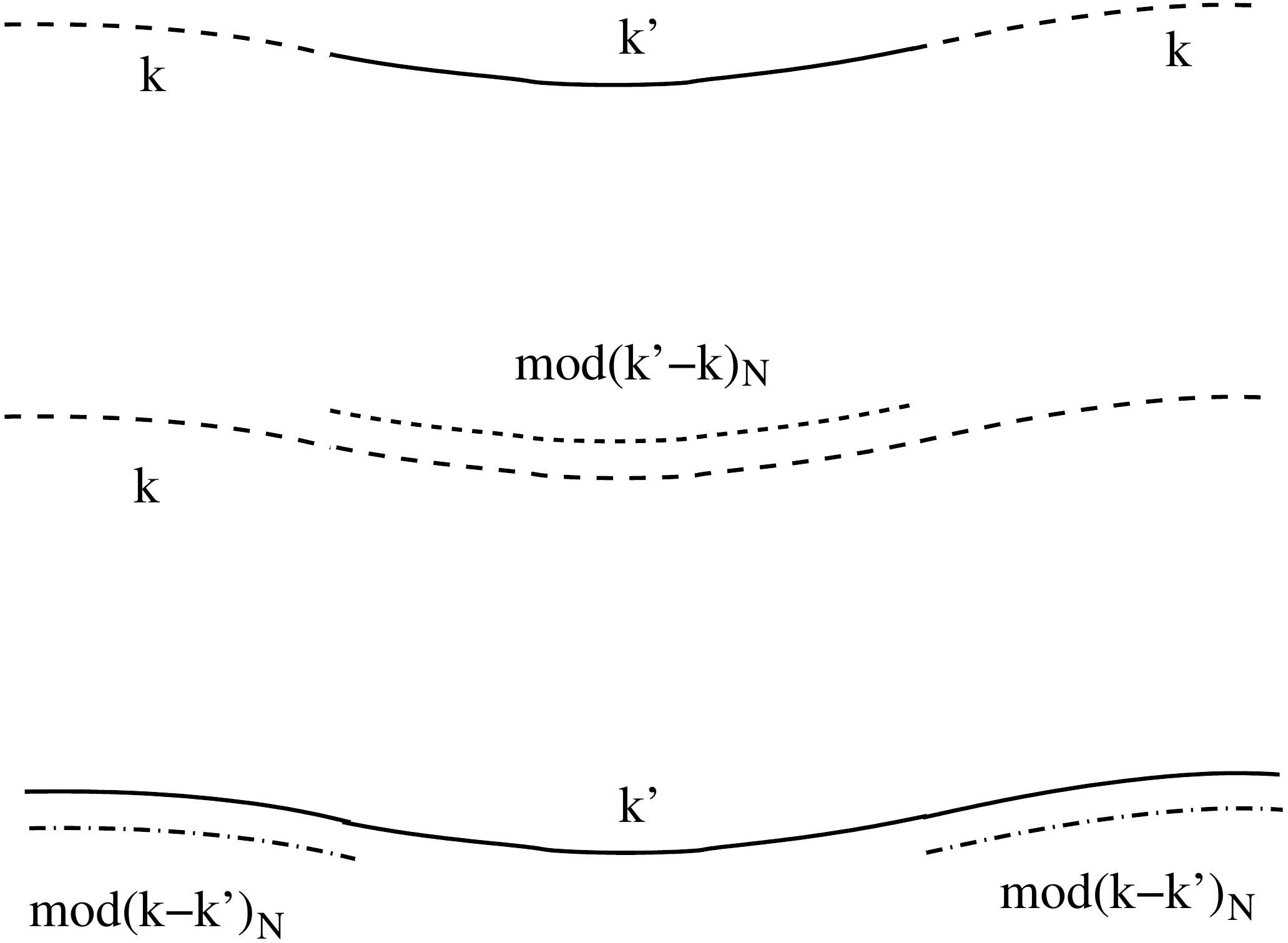}
\end{center}
\caption{Illustration of the ambiguity in labeling the ${\mathbb{Z}}_N$ 
topological sectors in presence of open vortices. Should the given 
configuration (top) be counted to the 
$k^{\mathrm{th}}$ (middle) or to the $k'^{\mathrm{th}}$ (bottom) sector?}
\label{fig:ambiguity}
\end{figure}
The presence of open vortices immediately blurs the distinction among them: 
does the state pictured at the top of Fig.~\ref{fig:ambiguity} belong to the 
$k^{\mathrm{th}}$ sector, resulting 
from the superposition of $k$ closed vortices with 
${\mathrm{mod}}(k'-k)_N$ open ones (middle picture), or does it belong to the 
$k'^{\mathrm{th}}$ sector, coming from the superposition of 
$k'$ closed vortices with ${\mathrm{mod}}(k-k')_N$ open 
ones winding in the other direction (bottom picture)? Clearly, there is no way 
to distinguish between them and assign the 
configuration to $Z(k)$ rather than $Z(k')$ in Eq.~(\ref{eq:free_en}). In other
words, a Wilson loop will never know if the vortex piercing it 
to generate the area law for its expectation value 
\cite{'tHooft:1977hy,'tHooft:1979uj} is open or closed: a 
confinement criterion based on the vortex free energy $F$ and hence on the
't Hooft loop can only make sense if open vortices are absent at any
temperature.

In this paper we will investigate a topological order parameter, the
center flux ${{z}}$, for
the transition between phases characterized by the presence of open or 
closed $\mathbb{Z}_2$ vortices in
$SU(2)$ Yang-Mills theories at $T=0$, discretized through standard plaquette 
actions. We will show that, depending on the action, the dimensions and
the volume, the theory can be either in a topologically
ordered or disordered phase; such distinction will persist at finite $T$. 
In the disordered phase open vortices dominate the vacuum and $\mathbb{Z}_2$ 
topological sectors are ill defined; the Hilbert space of 
Yang-Mills theories cannot be classified by a super-selection as in 
Eq.~(\ref{eq:hilbert}). Such disordered phase is compatible
with the presence of fundamental fermions; the
ordered phase, on the other hand, should be the correct one 
when coupling $SU(2)$ with
adjoint fermions, a popular candidate for infrared conformal gauge theories.

Besides this (perhaps lengthy) introduction, the rest of the paper is 
organized as follows: Sec.~\ref{sec:setup} contains 
details on the lattice setup, observables and simulation techniques; in 
Sec.~\ref{sec:res} the main results will be presented; Sec.~\ref{sec:con} 
contains the conclusions and outlook. Preliminary results of this 
investigations have been presented in Refs.~\cite{Burgio:2007np,Burgio:2013iaa}.

\section{Setup}
\label{sec:setup}

\subsection{Action and Observables}
\label{subsec:AO}

We will consider the $SU(2)$ mixed fundamental-adjoint Wilson action with 
periodic boundary conditions in 
$2 \leq d \leq 4$ Euclidean dimensions, as given in Eqs.~(\ref{eq:adj_action}, 
\ref{eq:fund_action}):
\begin{eqnarray}
S &=& \beta_{A}\sum_{P}\left(1-\frac{1}{3}\Tr_{A}\,U_{P}\right)+\beta_{F}\sum_{P}
\left(1-\frac{1}{2}\Tr_{F}\,U_{P}\right)
\nonumber\\
\frac{1}{a^{4-d}\,g^2} &=& \frac{1}{4}\beta_F+\frac{2}{3}\beta_A\,,
\label{eq:mixed}
\end{eqnarray}
where $a$ is the (dimensionful) lattice spacing 
and $U_P$ denotes the $1\times 1$ plaquette; all results can be easily
generalized to different boundary conditions. For higher groups $SU(N)$ the 
general picture 
should not change dramatically \cite{Creutz:1981qr,Creutz:1982ga,%
Drouffe:1982fe,Ardill:1982nk,Ardill:1982gm,Barresi:2006gq}. However, 
other representations than just the fundamental and adjoint are allowed. Many 
details might therefore depend on $N$; direct 
investigations of at least the $SU(3)$ case would be welcome.

In $d\geq 3$, $\mathbb{Z}_2$ monopoles can be defined for each elementary 
cube $c$ through the product:
\begin{equation}
\sigma_c = \prod_{{P}\in\partial c}\mathrm{sign}
(\Tr_{F} \,U_{{P}})\,,
\label{eq:mon}
\end{equation}
over all plaquettes $U_{{P}}$ belonging to its 
surface $\partial c$ \cite{Halliday:1981te,Halliday:1981tm,deForcrand:2002vs,%
Barresi:2003jq,Barresi:2006gq}. Notice how rescaling any link by a 
$\mathbb{Z}_2$ factor will leave $\sigma_c$ unchanged. 

\begin{figure}
\begin{center}
\includegraphics[width=0.7\textwidth]{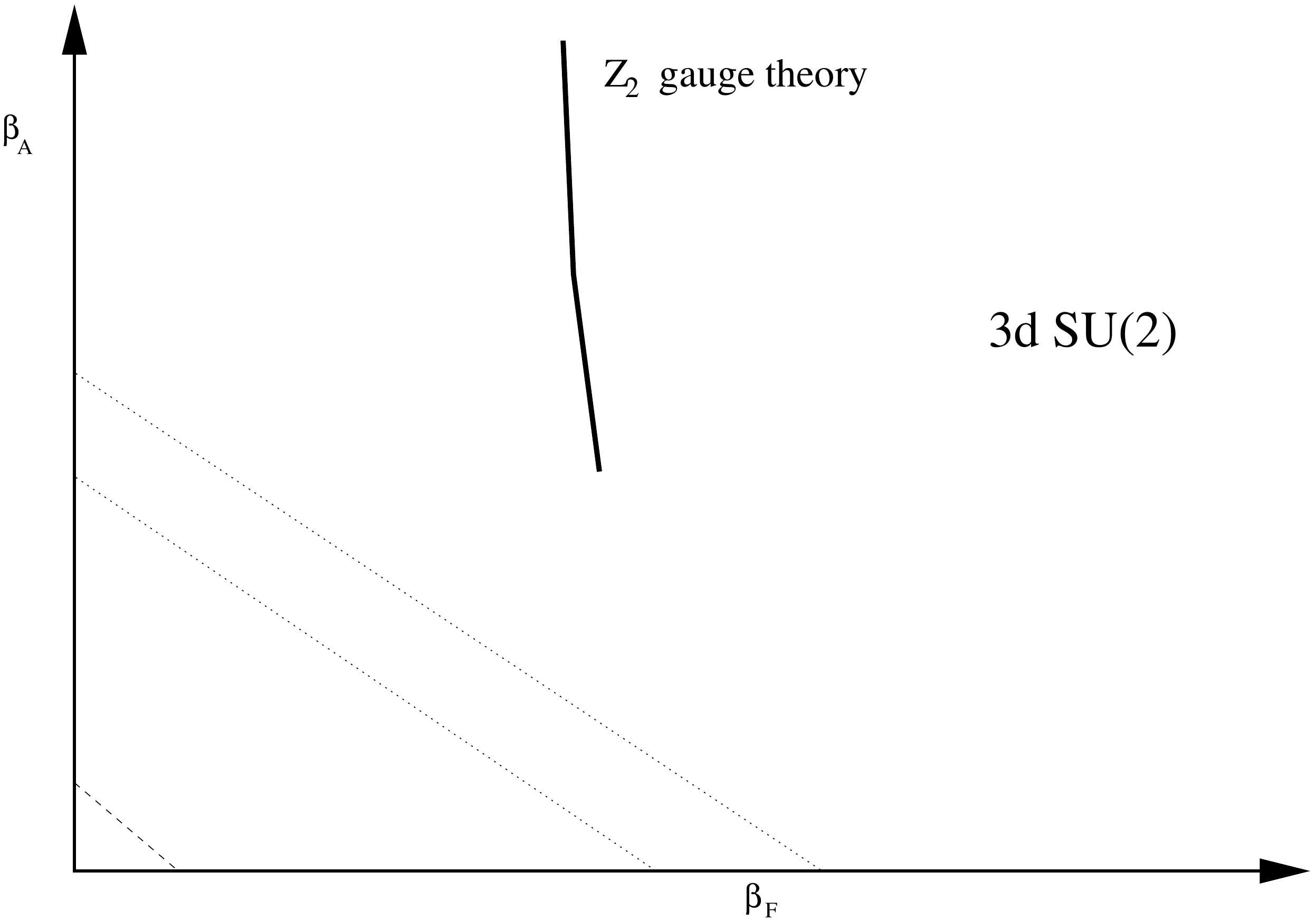}
\end{center}
\caption{Phase diagram of the fundamental-adjoint plane for $d=3$. 
Continuous lines indicate bulk transitions,
dashed lines the roughening transition, dotted lines the
crossover regions associated with $\mathbb{Z}_2$ monopoles. Similar diagrams 
hold for higher $N$.}
\label{fig:plane3}
\end{figure}

\begin{figure}
\begin{center}
\includegraphics[width=0.7\textwidth]{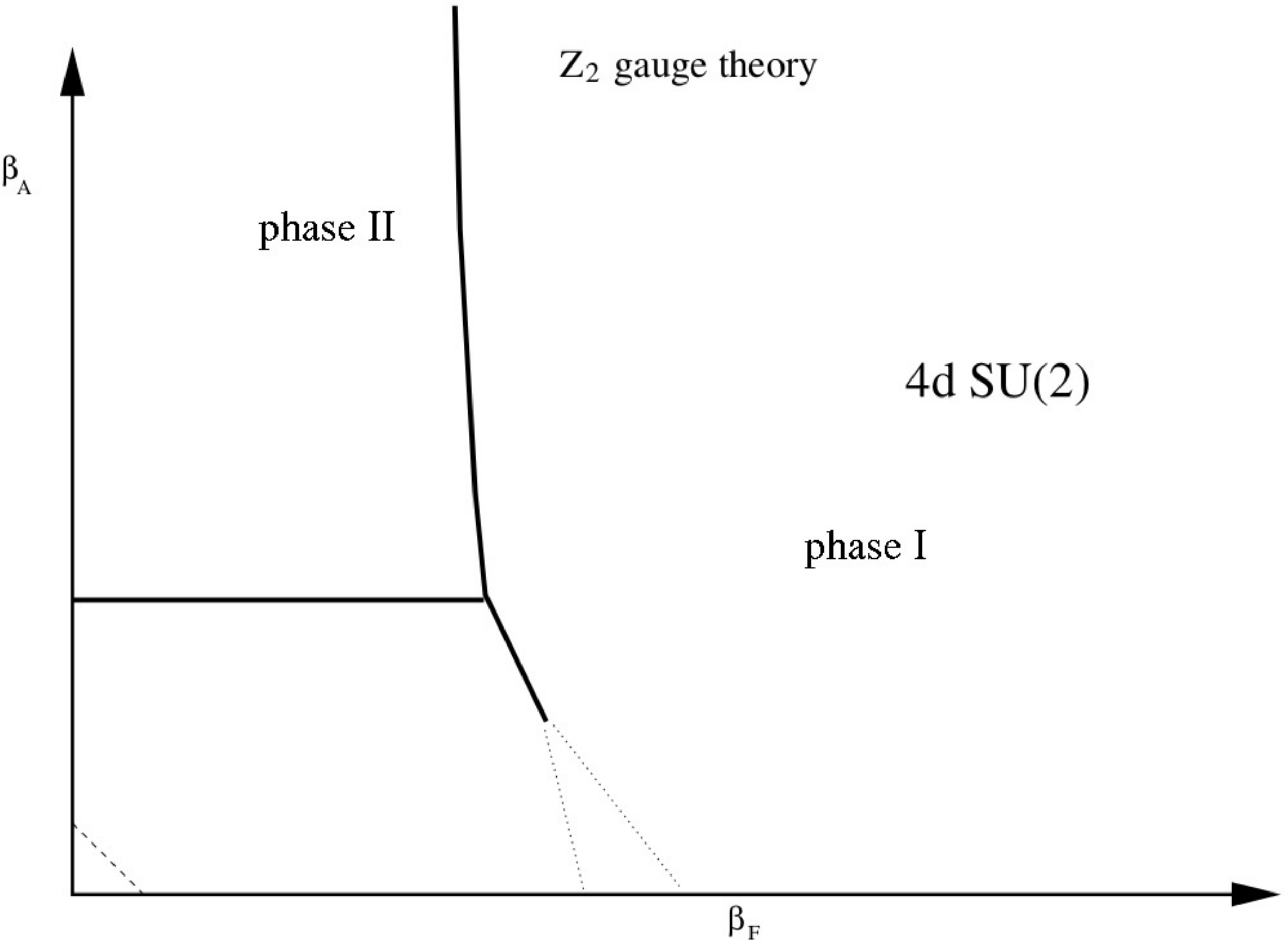}
\end{center}
\caption{Phase diagram of the fundamental-adjoint plane for $d=4$. 
Continuous lines indicate bulk transitions,
dashed lines the roughening transition, dotted lines the
crossover regions associated with $\mathbb{Z}_2$ monopoles. Similar diagrams 
hold for higher $N$.}
\label{fig:plane4}
\end{figure}

The $\mathbb{Z}_2$ monopole density should vanish in the continuum limit 
$g^2 \to 0$. This happens, however, in different ways, depending on the 
dimensions $d$ or the direction along which such limit is taken in the 
$\beta_F - \beta_A$ plane, and has been the subject of intense
investigations in the pioneering years of lattice gauge theories 
\cite{Drouffe:1980dp,Greensite:1981hw,Bhanot:1981eb,Halliday:1981te,%
Creutz:1981qr,Creutz:1982ga,Drouffe:1982fe,Ardill:1982nk,Ardill:1982gm,%
Baig:1986nn,Baig:1987qa,Baig:1987af,Bursa:2005tk}. For the 
$SU(2) - SO(3)$ case considered here the resulting phase diagrams in $d=3$ 
and 4 are sketched in Figs.~\ref{fig:plane3}, \ref{fig:plane4};
similar ones have been established for $N\geq 3$, see e.g. 
Refs.~\cite{Creutz:1981qr,Creutz:1982ga,Drouffe:1982fe,Ardill:1982nk,%
Ardill:1982gm}. Continuous lines indicate bulk transitions 
\cite{Greensite:1981hw,%
Bhanot:1981eb,Halliday:1981te,Baig:1987qa}, dashed lines the roughening 
transition \cite{Drouffe:1980dp}, dotted lines the crossover regions 
associated with $\mathbb{Z}_2$ monopoles \cite{Halliday:1981te,Baig:1987qa}. 
The vertical bulk transition line coming down form $\beta_A = \infty$ 
corresponds to the underlying $\mathbb{Z}_2$ gauge theory: in 
$d=3$ it ends at a finite point \cite{Baig:1987qa,Baig:1987af,Bursa:2005tk}, 
while in $d=4$ it joins the bulk transition line associated with 
$\mathbb{Z}_2$ monopoles
\cite{Greensite:1981hw,Bhanot:1981eb,Halliday:1981te}; from the endpoint of 
the latter a crossover region starts, extending beyond the $\beta_F$ axis.
In $d=2$, $\mathbb{Z}_2$ monopoles are of course absent. Furthermore, 
the $\mathbb{Z}_2$ gauge theory has no phase transition; 
apart from the roughening transition \cite{Drouffe:1980dp}, the corresponding 
phase diagram 
should therefore be free of any bulk effects, including crossovers. 

From Fig.~\ref{fig:plane4} it is obvious that two distinct continuum
limits in $d=4$ exist, depending if $g^2 \to 0$ in Eq.~(\ref{eq:mixed}) is 
taken within Phase I or II. Phase II at fixed twist has been shown 
in Refs.~\cite{deForcrand:2002vs,Barresi:2006gq} to be equivalent to a 
positive plaquette model \cite{Mack:1981gy,Bornyakov:1991gq,Fingberg:1994ut}
with fixed twisted boundary conditions. Although such model and the fundamental
Wilson action seem to describe the same physics, the two phases are always
separated by a bulk transition line.\footnote{The authors of
 Ref.~\cite{deForcrand:2002vs} also proved that the
1$^{\mathrm{st}}$ order line separating the two phases is just a finite volume 
effect: at high enough volume Phase I and II will be always separated 
by a 2$^{\mathrm{nd}}$ order line.} 
What is thus the difference, if any, between them? 

A first hint towards an explanation to this (long neglected) puzzle is 
given by the results of Refs.~\cite{Burgio:2006dc,%
Burgio:2006xj,Barresi:2001dt,Barresi:2002un,Barresi:2003jq,Barresi:2003yb,%
Barresi:2004qa,Barresi:2004gk,Burgio:2005xe,Barresi:2006gq}:
in the continuum limit the $d=4$ adjoint theory ($\beta_F \equiv 0$),
which lies precisely within phase II, possesses well defined $\mathbb{Z}_2$ 
topological sectors, i.e. no open vortices: the Hilbert-space decomposition
defined in Eq.~(\ref{eq:hilbert}) works! On the other hand, one can easily 
check 
that in phase I, across all crossovers, the $\mathbb{Z}_2$ monopole density 
vanishes quite slowly as $g^2 \to 0$: their persistence in the
weak coupling phase should reflect itself in the presence of
open $\mathbb{Z}_2$ vortices, possibly spoiling Eq.~(\ref{eq:hilbert}). 
Could the difference between phase I and II lie in whether such super-selection 
rule is indeed realized for the Hilbert space of Yang-Mills theories?
To find out, we can start from the twist operator, which ``counts'' the 
$\mathbb{Z}_2$ vortices piercing all parallel planes for a fixed choice of 
$\mu$-$\nu$
\cite{'tHooft:1979uj,deForcrand:2002vs}:
\begin{equation}
z_{\mu\nu} = \frac{1}{L^{d-2}}\,\sum_{\hat{y}\,\bot\,{\mu\nu}-{\rm plane}}\; 
\prod_{\hat{x}\, \in \,{\mu\nu}-{\rm plane}} 
\mathrm{sign}({\Tr}_{F}U_{\mu\nu}(\hat{x},\hat{y}))\,.
\label{eq:twist}
\end{equation}
$U_{\mu\nu}$ and $\hat{x}$, respectively, denote a $1\times 1$ plaquette 
and point lying 
in the $\mu$-$\nu$ plane, while $\hat{y}$ denotes a point on its co-set,
which is obviously empty in $d=2$; only a single plane contributes to 
the sum in this case. Notice how $z_{\mu\nu}$, like $\sigma_c$, 
is unaffected by any multiplication of links by a center element, i.e.
it is insensitive to the spurious $\mathbb{Z}_2$ gauge degrees of freedom.

If topological sectors are well defined, all parallel 
planes will contribute with the same sign to the sum in 
Eq.~(\ref{eq:twist}). For any fixed $\mu$ and $\nu$, $z_{\mu\nu}$ can thus only 
take the values $\pm 1$, depending 
on the boundary conditions chosen.\footnote{Only for $\beta_F =0$, i.e. 
along the $\beta_A$ axis, the $z_{\mu\nu}$ are allowed to tunnel among different 
topological sectors, provided that an ergodic algorithm capable of 
overcoming the large barriers among them is used. In this case the $z_{\mu\nu}$ 
can take both values $\pm 1$ \cite{Burgio:2005xe,Burgio:2006dc}.} 
E.g., for the periodic boundary conditions considered in this paper, the 
topological sector must always be trivial: $z_{\mu\nu} \equiv 1$ 
$\forall \mu, \nu$. 
When, however, topological sectors are ill defined the contributions to the 
sum in Eq.~(\ref{eq:twist}) can change from 
plane to plane; in particular, 
if open $\mathbb{Z}_2$ 
vortices pierce the planes randomly, all $z_{\mu\nu}$ will 
average to zero. To make such statement quantitative and
characterize how the transition from the disordered
to the ordered regime takes place we define a (non-local!) order 
parameter, the center flux ${{z}}$, such that its expectation
value $\langle {{z}}\rangle \equiv 1$ if, whatever the boundary conditions, 
vortex topology takes the correct value expected from the super-selection rule, 
while $\langle {{z}}\rangle \equiv 0$ when $\mathbb{Z}_2$ fluxes are maximally 
randomized. For $d\geq 3$:
\begin{equation}
{{z}} =\frac{2}{d(d-1)} \sum_{\mu>\nu=1}^{d} |z_{\mu\nu}|\,,
\label{eq:op}
\end{equation}
while for $d=2$, since $|z_{12}| \equiv 1$, we will define:
\begin{equation}
{{z}} = 1-|z_{12} - \langle z_{12} \rangle|\,.
\label{eq:op2_al}
\end{equation}
Notice that the latter definition will only work as long as $\beta_F \neq 0$, 
i.e. when the 
$d=2$ theory cannot tunnel among topological sectors.\footnote{The definition 
of the center flux in $d=2$ might also be adjusted to the pure adjoint theory 
as long as 
no ergodic algorithm is available in the ordered phase. The 
issue is similar to that encountered for e.g. an Ising model when simulating 
the low-temperature phase with a cluster algorithm.}
In the following we will investigate, either analytically (in $d=2$) or via 
Monte-Carlo simulations (for $d\geq 3$), the behaviour of the center flux and 
its susceptibility:\footnote{Since ${{z}}$ is non-local,
one could argue that the volume factor 
should be substituted by the number of planes $\frac{d (d-1)}{2}
L^{d-2}$. This would however just change the critical exponent for
$\chi_{{z}}$ from $L^\gamma \to L^{\gamma-2}$, which could be 
re-absorbed in the definition of the hyper-scaling relations. Moreover
for each plane up to 
$L^2$ vortices can form, summing up again to 
$L^d$. To underline the 
analogies of our results with the Kosterlitz-Thouless literature 
we will thus stick to the standard definition. Anyhow, 
critical behaviours are controlled by a diverging correlation length
$\xi$, which remains unaffected by any re-scaling of $\chi_{{z}}$. 
}
\begin{equation}
\chi_{{z}} = L^d\,\;\left({{z}}-\langle {{z}} \rangle\right)^2\,.
\label{eq:susc}
\end{equation}

\subsection{Algorithm}
\label{sec:alg}

Simulations for $\beta_A =0$, i.e. along the $\beta_F$ axis, have been 
performed using a standard heat-bath algorithm followed by 
micro-canonical steps. Although this cannot be extended to 
$\beta_A \neq 0$, as long as also $\beta_F \neq 0$ one can use the 
biased Metropolis $+$ micro-canonical algorithm introduced in 
Refs.~\cite{Bazavov:2005zy,Bazavov:2005vr}.\footnote{See e.g. 
Ref.~\cite{Lucini:2013wsa} for a recent application. A similar 
algorithm had been proposed in Ref.~\cite{Hasenbusch:2004yq} for $SU(3)$.} 
The lookup tables for the pseudo-heat-bath probability need to be fixed 
beforehand: sizes between $32 \times 32$ and $64 \times 64$ were found to be 
sufficient \cite{Bazavov:2005zy,Bazavov:2005vr}. As long as 
$\beta_F \gg \beta_A$, the algorithm is for all practical purposes just 
as efficient as an heat-bath, as the amount of accepted proposals stays
well above $95 \%$. On the other hand, whenever 
$\beta_F \ll \beta_A$ the rejected pseudo-heat-bath and micro-canonical 
updates increase considerably. 
This becomes a real issue when simulating around the peaks of the 
susceptibility 
Eq.~(\ref{eq:susc}), where auto-correlations for ${{z}}$ and 
$\chi_{{z}}$ become quite large.\footnote{Other observables remain, 
on the other hand, mostly unaffected.} 
One can try to combat such critical slowing down,\footnote{The critical 
slowing down
appears of course also in the limit $g^2\to 0$, i.e. for large $\beta_{F}$ 
and/or $\beta_{A}$.}
unavoidable when dealing 
with any phase transition, by increasing the number of micro-canonical
steps per biased Metropolis update. Unfortunately this turns out to be less 
efficient than for the $\beta_F \gtrsim \beta_A$ case or for the heat-bath 
algorithm; only with runs of order $\sim 10^8$ sweeps one eventually reaches 
a good signal-to-noise ratio for $\chi_{{z}}$. Since the $d=3$ case
will anyway turn out to be the most interesting from the point of view of the 
critical behaviour,
while in $d=2$ analytic results allow to gain otherwise control of the problem, 
we have limited 
a precise finite size scaling (FSS) \cite{Fisher:1972zza} 
analysis to determine the properties of the 
transition to the $\beta_A=0$, $d=3$ case. Still, we have performed simulations
for a whole range of parameters and lattice sizes $L$ in $2 \leq d \leq 4$, 
trying to explore the whole $\beta_F - \beta_A$ plane. 
We have nevertheless avoided phase II of the $d=4$ phase diagram in 
Fig.~\ref{fig:plane4}, since it would have
called for completely different simulation techniques; see 
Refs.~\cite{Burgio:2006dc,%
Burgio:2006xj,Barresi:2001dt,Barresi:2002un,Barresi:2003jq,Barresi:2003yb,%
Barresi:2004qa,Barresi:2004gk,Burgio:2005xe,Barresi:2006gq} for results in
this parameter region.

\section{Results}
\label{sec:res}

\subsection{$d=2$}

The $SU(2)$ theory in $d=2$ offers the chance to tackle our problem 
analytically 
\cite{Eriksson:1980rq,Lang:1980sz}.
The probability distribution for the mixed action in 
Eq.~(\ref{eq:mixed}) reads:
\begin{equation}
d\, \rho(\theta,\beta_F,\beta_A) \propto d\,\theta \sin^2{\theta} \,
\mathrm{e}^{\,\beta_F\,\cos{\,\theta} \,+\, \frac{4}{3}\,\beta_A \cos^2{\theta}}\,,
\label{eq:prob0}
\end{equation}
so that
the probability for a plaquette to have negative trace is simply given by: 
\begin{equation}
p(\beta_F,\beta_A) = \frac{\int_{\frac{\pi}{2}}^\pi d\,\theta \sin^2{\theta} \,
\mathrm{e}^{\,\beta_F\,\cos{\,\theta} \,+\, \frac{4}{3}\,\beta_A \cos^2{\theta}}}{\int_0^\pi 
d\,\theta\, \sin^2{\theta} \,
\mathrm{e}^{\,\beta_F\,\cos{\,\theta} \,+\, \frac{4}{3}\,\beta_A \cos^2{\theta}}}\,.
\label{eq:prob}
\end{equation}
The limiting cases $\beta_{F,A} \to 0\,, \infty$ 
can be carried out explicitly, giving:
\begin{eqnarray}
p(\beta_F \equiv 0,\beta_A)\; &=& \frac{1}{2}
\nonumber\\
p(\beta_F \equiv \infty,\beta_A) &=&\, 0 \label{eq:p_A}\\
p(\beta_F,\beta_A \equiv 0)\; &=& \frac{1}{2}\left[1-
\frac{L_1(\beta_F)}{I_1(\beta_F)}\right]
\nonumber\\
p(\beta_F,\beta_A \equiv \infty) &=&\; \frac{1}{1 + \mathrm{e}^{2\,\beta_F}}\,,
\label{eq:prob_lim}
\end{eqnarray}
where $L$ and $I$ denote the modified Struve and Bessel functions, 
respectively \cite{abramowitz+stegun}.

\begin{figure}
\begin{center}
\includegraphics[width=0.7\textwidth]{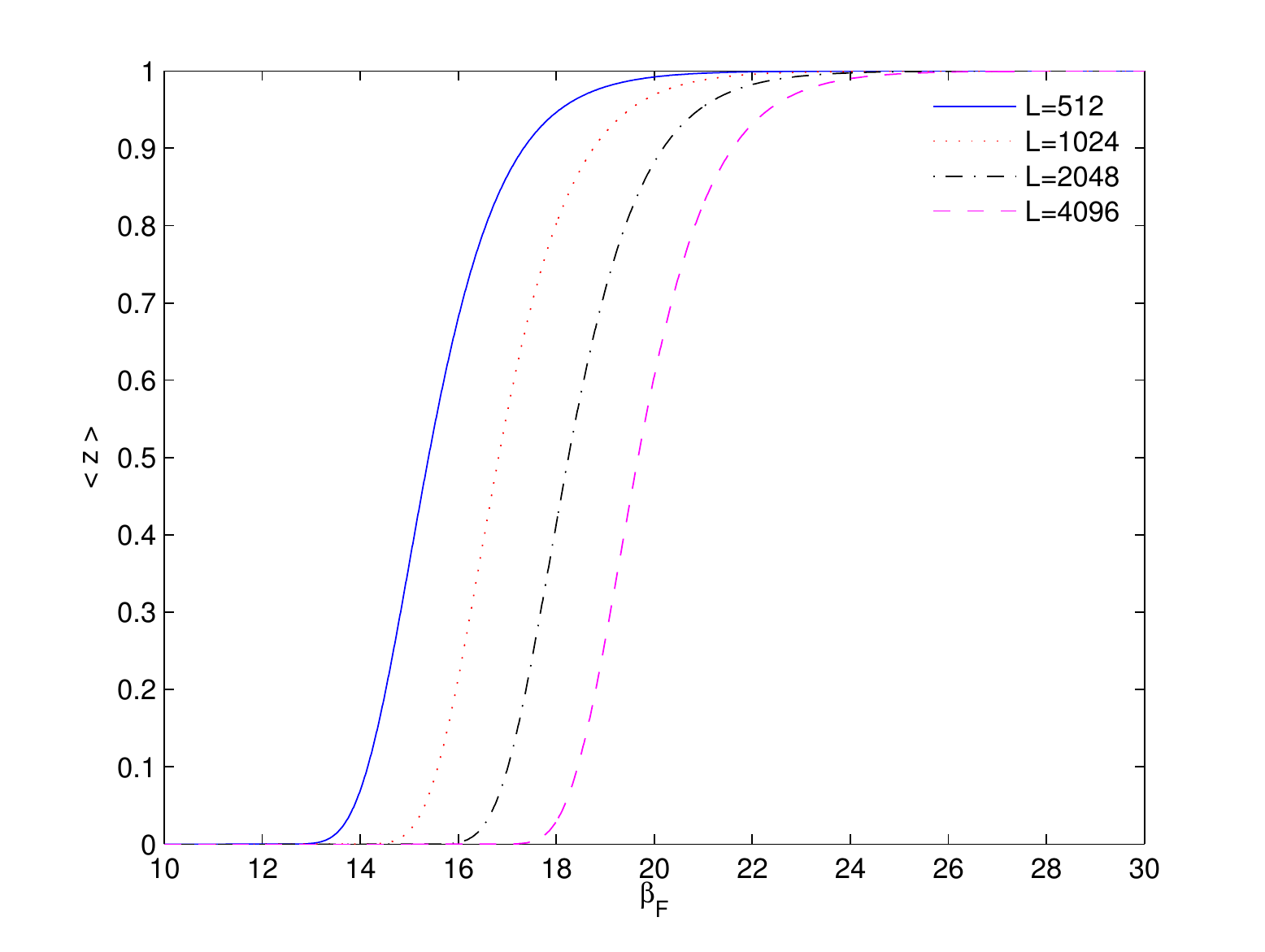}
\end{center}
\caption{Order parameter $\langle {{{z}}} \rangle$ in $d=2$ 
along $\beta_F$ for $L=512$, 1024, 2048 and 4096.}
\label{fig_1a}
\end{figure}

\begin{figure}
\begin{center}
\includegraphics[width=0.7\textwidth]{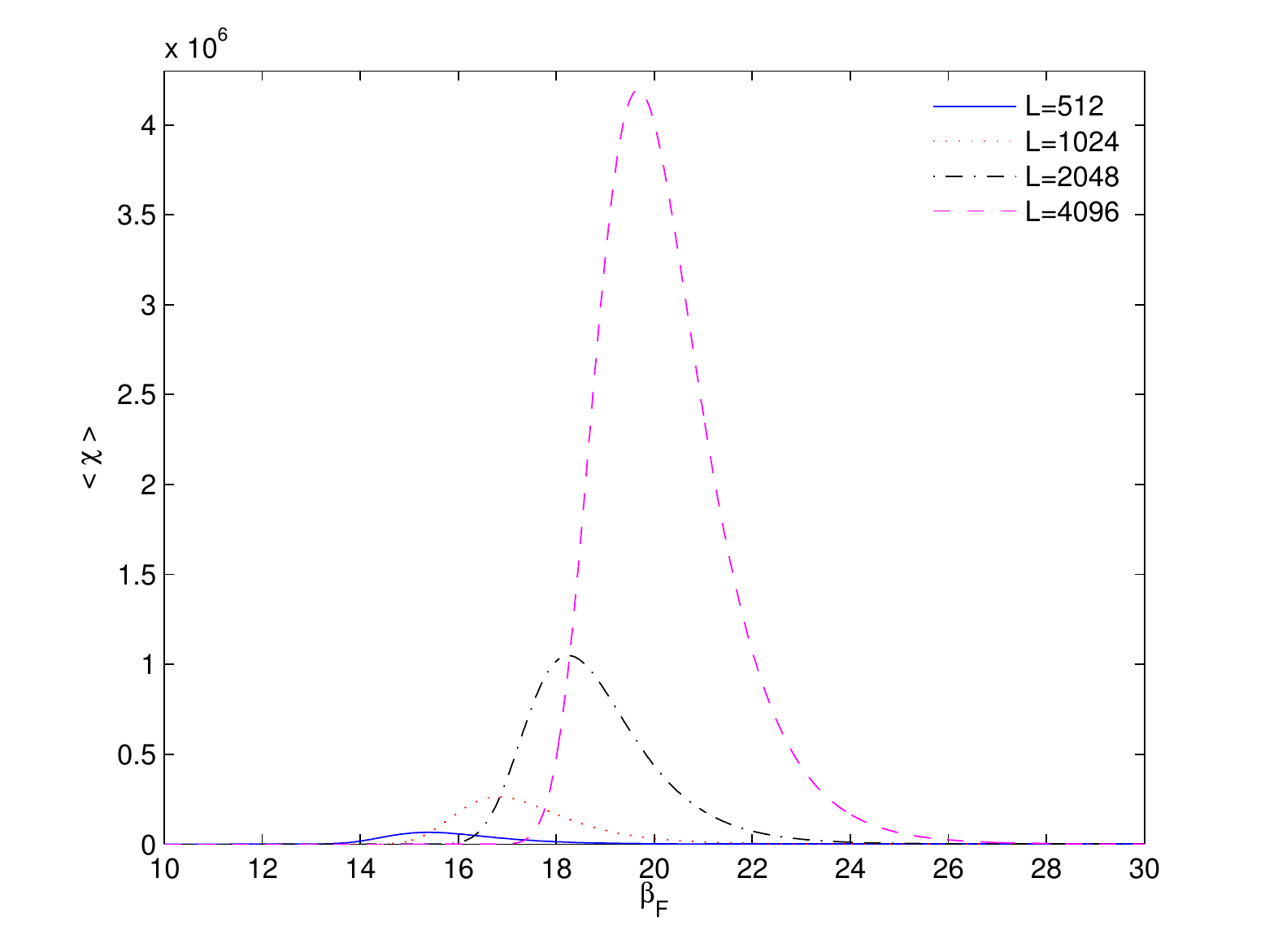}
\end{center}
\caption{Susceptibility $\chi_{{z}}$ in $d=2$ 
along $\beta_F$ for $L=512$, 1024, 2048 and 4096.}
\label{fig_1b}
\end{figure}

\begin{figure}
\begin{center}
\includegraphics[width=0.7\textwidth]{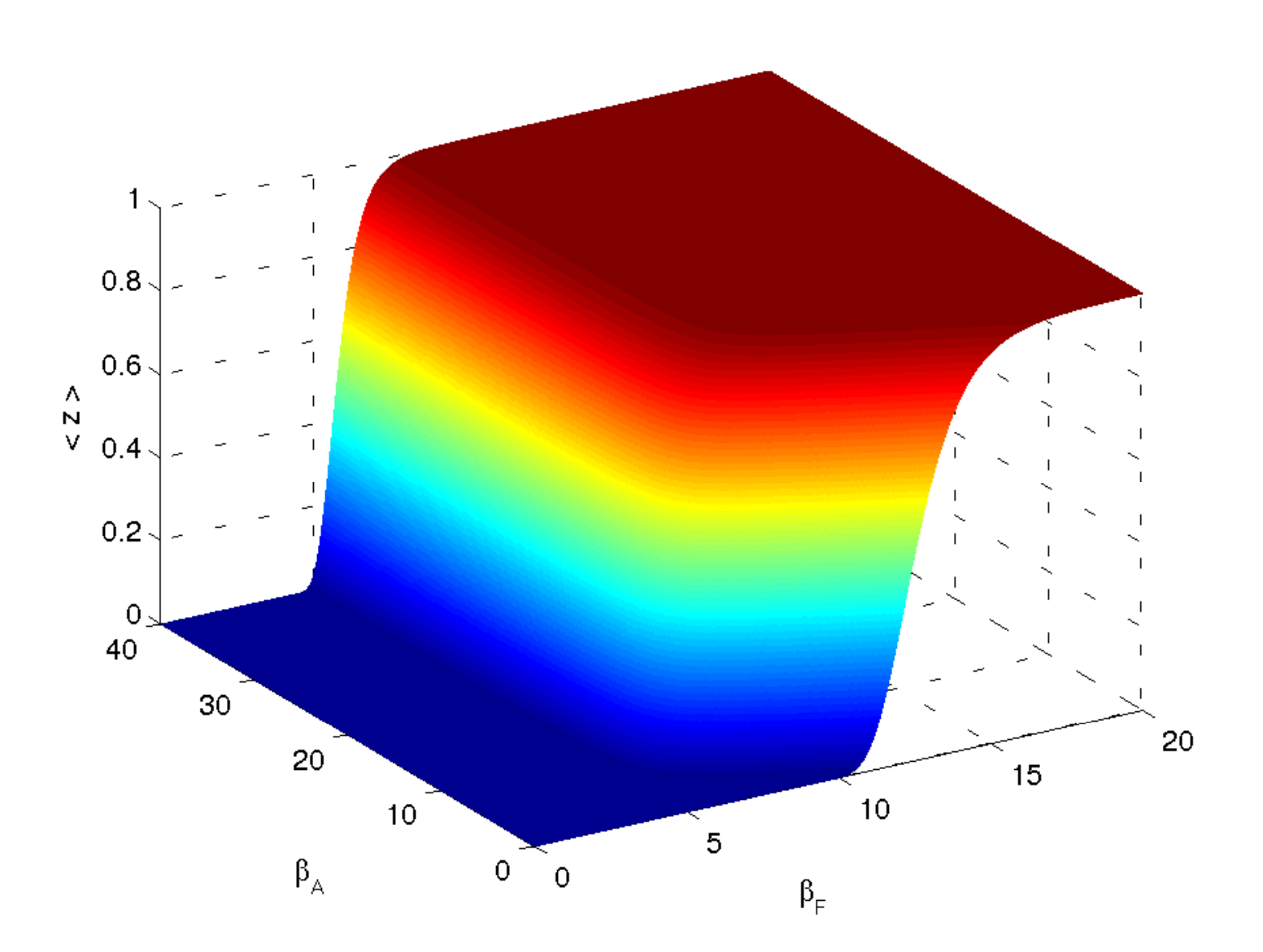}
\end{center}
\caption{Order parameter $\langle {{{z}}} \rangle$ in $d=2$ for 
fixed $L=128$.}
\label{fig_2}
\end{figure}

\begin{figure}
\begin{center}
\includegraphics[width=0.7\textwidth]{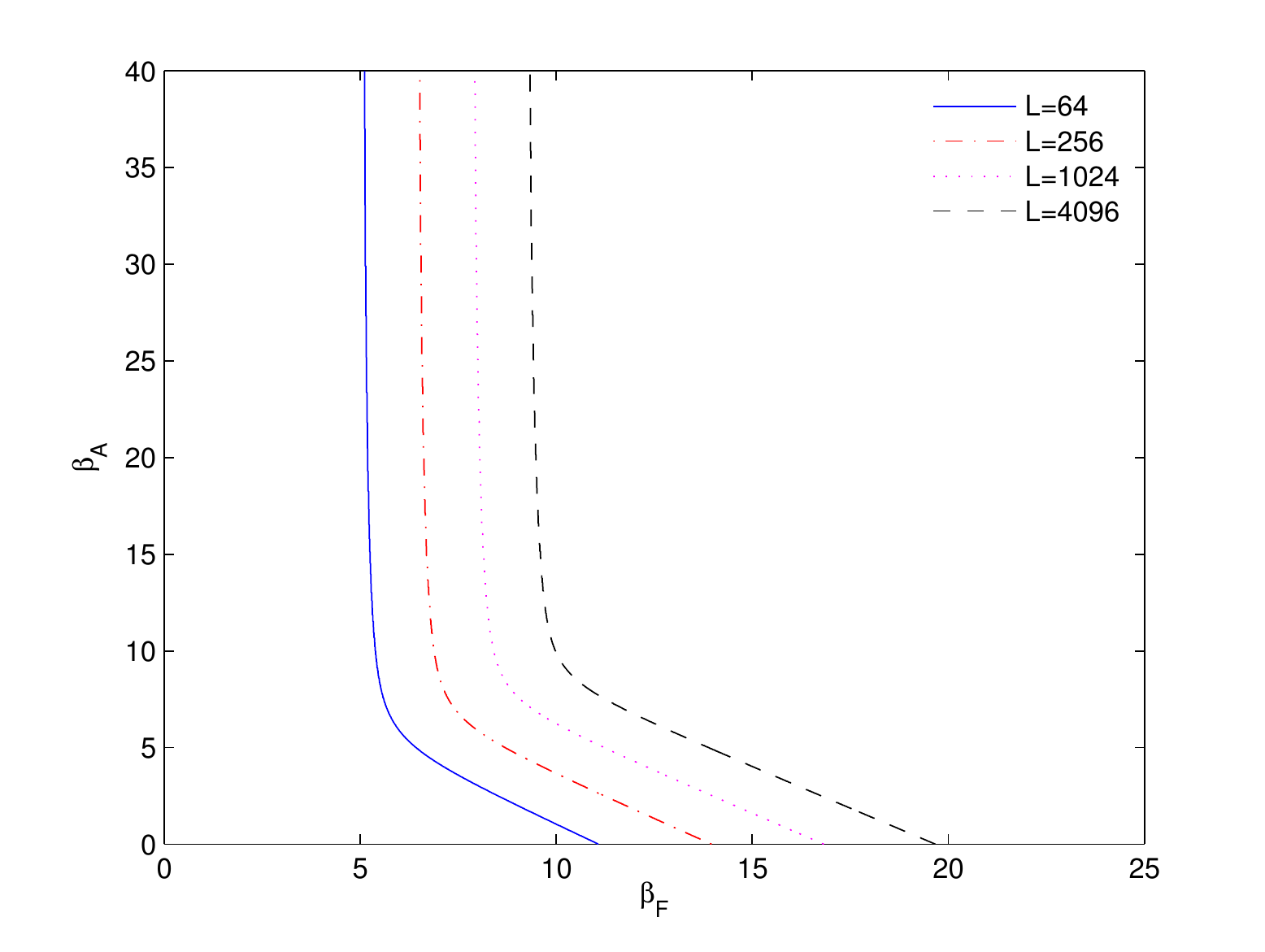}
\end{center}
\caption{Peak curves of the susceptibility $\chi_{{z}}$ in the 
$\beta_F - \beta_A $ plane for $L=64$, 256, 1024 and 4096.}
\label{fig_2a}
\end{figure}

For fixed volume $L^2$ the order parameter ${{z}}$ and its 
susceptibility $\chi_{{z}}$ 
are given by \cite{Bursa:PC}:\footnote{We are indebted 
to F. Bursa for 
precious correspondence on the derivation of the above expressions for ${{z}}$ 
and $\chi_{{z}}$. Just to be on the safe side, we have also cross-checked all 
analytic results with Monte-Carlo simulations up to $L=1024$; these become of 
course inefficient as $\beta_A$ gets large...}
\begin{eqnarray}
\langle {{{z}}} \rangle &=& \mathrm{e}^{-4 L^2 p(\beta_F,\beta_A)}
\nonumber\\
\langle\chi_{{z}}\rangle &=& L^2 \left[\mathrm{e}^{-4 L^2 p(\beta_F,\beta_A)}-
\mathrm{e}^{-8 L^2 p(\beta_F,\beta_A)}\right]\,.
\label{eq:chi_L}
\end{eqnarray}
The above expressions are plotted, for $\beta_A = 0$, in Fig.~\ref{fig_1a} 
and \ref{fig_1b}; a similar behaviour extends to the whole 
$(\beta_F,\beta_A)$ plane, see Fig.~\ref{fig_2}, where the center flux is 
plotted for fixed $L=128$. 
We can clearly distinguish a low $\beta_F$, ``strong''
coupling regime, where $\langle {{z}} \rangle =0$ and the topology is 
ill defined, 
from a high $\beta_F$, ``weak'' coupling one, where 
$\langle {{z}} \rangle =1$, 
the correct value it should have if the vacuum satisfies Eq.~(\ref{eq:vac}). 
For higher $L$ the transition ``front'' simply moves to the right, i.e. 
higher $\beta_F$;
see Fig.~\ref{fig_2a}, where the curves along which the 
susceptibility $\chi_{{z}}$ peaks are plotted for $L=64$, 
256, 1024 and 4096.

As usual in a FSS analysis, we can determine the properties of 
the transition by defining the pseudo-critical couplings $(\beta^c_F(L), 
\beta^c_A(L))$ at finite $L$ as those for which the correlation length 
$\xi \simeq L$ \cite{Fisher:1972zza}. These can be 
identified through the peaks of the susceptibility $\chi_{{z}}$ 
(see Fig.~\ref{fig_2a}); since $p(\beta_F,\beta_A)$
has no stationary points, from Eq.~(\ref{eq:chi_L}) one simply needs to solve:
\begin{equation}
p(\beta^c_F,\beta^c_A) = \frac{\log{2}}{4\, L^2}\,.
\label{eq:max_p}
\end{equation}
Substituting the above value into Eq.~(\ref{eq:chi_L}) we get for the scaling 
of the center flux and its susceptibility with $L$:
\begin{eqnarray}
{{z}}(\beta^c_F,\beta^c_A) &=& \frac{1}{2}
\nonumber\\
\chi_{{z}}(\beta^c_F,\beta^c_A) &=& \frac{L^2}{4}\,.
\label{eq:max_chi}
\end{eqnarray}
As for the scaling of the pseudo-critical points with $L$, from 
Eqs.~(\ref{eq:p_A}) we have that along lines parallel to the $\beta_A$ axis 
$\langle {{z}} \rangle = 0$. Otherwise, we can fix a line 
$\beta_A = f(\beta_F)$ and solve: 
\begin{equation}
p(\beta_F,f(\beta_F)) = \frac{\log{2}}{4\, L^2}
\label{eq:fss_beta}
\end{equation}
for $\beta_F$. In particular, from Eq.~(\ref{eq:prob_lim}) and using
the asymptotic expansion \cite{abramowitz+stegun}:
\begin{equation}
p(\beta_F,\beta_A \equiv 0) \sim
\sqrt{\frac{2 \beta_F}{\pi}}\mathrm{e}^{-\beta_F}\,\left[1+
{\cal{O}}(\frac{1}{\beta_F})\right]
\end{equation}
we get for the two limiting cases $\beta_A \to 0\,, \infty$:
\begin{eqnarray}
\beta^c_F(L)|_{\beta_A = 0} &\sim& \log{L^2} + \frac{1}{2}\log{\log{L^2}} + {\cal{O}}(1)
\label{eq:ffs_2d_F}
\\
\beta^c_F(L)|_{\beta_A = \infty} &\sim& \frac{1}{2}\log{L^2} + {\cal{O}}(1)\,:
\label{eq:fss_2d_A}
\end{eqnarray}
shifting Eqs.~(\ref{eq:chi_L}) by Eqs.~(\ref{eq:ffs_2d_F}, \ref{eq:fss_2d_A}) 
$\langle {{z}} \rangle$ and $\langle \chi_{{z}} \rangle$ will fall on 
top of each other.

Inverting Eqs.~(\ref{eq:ffs_2d_F}, \ref{eq:fss_2d_A}), one can
extract the critical behaviour of the correlation length $\xi \sim L$ for 
$\beta_A \to 0\,, \infty$, where the pre-factors come from 
the ${\cal{O}}(1)$ terms:
\begin{eqnarray}
\xi|_{\beta_A = 0} \;&\sim& \sqrt[4]{\frac{\pi \log^2{2}}{2 \beta_F}}\;\cdot\, 
\mathrm{e}^{\frac{1}{2}\beta_F}
\label{eq:ess0}\\
\xi|_{\beta_A = \infty} &\sim& \sqrt{\frac{\log{2}}{4}}\;\cdot\, \mathrm{e}^{\beta_F}\,.
\label{eq_ess}
\end{eqnarray}
Similar expressions will hold for any direction $\beta_A = f(\beta_F)$ along 
which the continuum limit $\beta_F \to \infty$ is taken.\footnote{Of course, 
this only holds as long as $f(\beta_F) \sim \beta_F$ for large $\beta_F$.} 
At $T>0$ one can simply substitute $L^2 \to L_s\cdot L_t$ in 
Eqs.~(\ref{eq:max_p}, \ref{eq:max_chi}, \ref{eq:ffs_2d_F}, \ref{eq:fss_2d_A}). The scaling behaviour remains
thus, up to a factor, unchanged when taking the thermodynamic limit 
$L_s\to\infty$: the critical behaviour will persist for any fixed $L_t$,
i.e. at any temperature.

Compare now the above scaling with the critical behaviour of the 
Kosterlitz-Thouless universality class 
\cite{Kenna:1996bs,Kosterlitz:1973xp,Kosterlitz:1974sm} as a function of the 
reduced coupling $\beta_{\mathrm{red}}$:
\begin{align}
\xi_{KT} \sim & \,K\,\mathrm{e}^{A \, \beta_{\mathrm{red}}^{\nu}}\\
\beta_{\mathrm{red}}^{-1} = &  \left|\beta^{-1}-\beta^{-1}_c\right|
\propto \left|T-T_c\right|\,.
\end{align}
Albeit with a different critical exponent, $\nu=1$ and $\nu = 1/2$ 
respectively, both cases show essential scaling, 
i.e. the correlation length diverges exponentially
as one approaches the critical coupling, which in our case is 
$\beta^c_F = \infty$. Mimicking now a well-known argument
\cite{Kosterlitz:1973xp,Kosterlitz:1974sm},
we can give a simple explanation for the behaviour found in 
Eqs.~(\ref{eq:ess0}, \ref{eq_ess}).
At weak coupling the free energy cost
to change the sign of a plaquette is $f \sim 2 \beta_F$; the density 
of negative plaquettes will thus be controlled by a Boltzmann factor 
$\rho \sim \exp{(- f)}$. On the
other hand the possible positions for this sign flip will scale like $L^2$ 
and the balance between free energy and entropy gives 
$L \sim \rho^{-1/2} = \exp{(\beta_f)} \simeq \xi$.\footnote{We wish to thank P. 
de Forcrand for useful comments about this point.} Up to the power correction
for $\beta_A=0$ case, Eq.~(\ref{eq:ess0}), this simple 
argument works quite well, contrary to the $XY$-model, where
it cannot explain renormalization effects leading to the non trivial
critical exponent $\nu = 1/2$. Moreover, since the minimal distance among
vortices can be reliably estimated with that along a plane intersecting them,
such picture should (roughly) hold in higher dimensions as well.

We could in principle explore the similarities with the Kosterlitz-Thouless 
transitions further. Although, as far as we know, for the $XY$-model no 
local order parameter is available, one can
couple the theory to an external magnetic field $h$ and study the 
analytical continuation of the partition function $Z(\beta,h)$ to the complex 
plane. (Hyper-) scaling relations will then hold among the critical exponents 
of $\xi$, of the magnetic susceptibility 
$\chi_h \sim \xi^{2-\eta} 
\log^{-2 r}{\xi}$, of the specific heat $C_s$ and of the edge of the 
Lee-Yang zeroes \cite{Kenna:1996bs}.
We will avoid such a throughout analysis in our case, for which a dedicated 
paper would be needed. Let us however just briefly comment on two points. 
First, from Eq.~(\ref{eq:prob0}) we can explicitly calculate the reduced 
partition function and the specific heat in our usual limiting cases:
\begin{align}
Z \underset{\beta_A\to 0}{\propto} &\; \frac{1}{\beta_F}\, I_1(\beta_F)
\label{eq:Zf}\\
Z \underset{\beta_A\to \infty}{\propto} &\, 
\frac{\mathrm{e}^{\beta_A}}{\sqrt{\beta_A^3}}\, \cosh{\beta_F}
\label{eq:Z}\\
C_s\underset{\beta_A\to 0}{=} & \;\frac{3}{2\, \beta^2_F}
\label{eq:Cs_f}\\
C_s \underset{\beta_A\to \infty}{=} &\; 1-\tanh^2{\beta_F}\,.
\label{eq:Cs_A}
\end{align}
Inserting Eqs.~(\ref{eq:ffs_2d_F}, \ref{eq:fss_2d_A}) into 
Eqs.~(\ref{eq:Cs_f}, \ref{eq:Cs_A}) and assuming that no other 
contribution besides the singular one exist \cite{Kenna:1996bs}, we see 
that the 
critical behaviour for $C_s$ should change (continuously?) from $\log^{-2}L$ to 
$L^{-2}$. Second, in our case we have direct access to a non local, topological 
order parameter, for which we can determine a critical exponent,
${{z}}\sim M\,L^{-\beta}$; from Eq.~(\ref{eq:max_chi}) we have $\beta = 0$. If we 
would like to study the extended partition function $Z(\beta_F,\beta_A,h)$ we 
could simply add a term $S_h = h\,{{z}}$ to the action. Although a direct 
calculation would go beyond the scope of this paper, it is obvious that for 
fixed $L$ a sufficient condition to align the center flux ${{z}}$ is realized 
if $\beta_F \to\infty$: the fundamental coupling plays the role of a "mock" 
$\mathbb{Z}_2$ magnetic field. Indeed, from Eqs.~(\ref{eq:Zf}, \ref{eq:Z}) the 
zeros of $Z$ in the complex $\beta_F$ plane all lie on the imaginary axis, in 
agreement with the Lee-Yang theorem \cite{Lee:1952ig}.

Let us finally turn to the continuum limit.
From the above discussion it is clear that taking the thermodynamic limit 
$L\to \infty$ before the weak coupling limit $\beta_F \to \infty$, as one 
should, i.e. taking the Euclidean volume 
$V=(a\,L)^2 \to \infty$ (or, at finite temperature, 
$V_s = (a\,L_s) \to \infty$), the theory remains stuck in the disordered
phase $\langle {{z}} \rangle = 0$: no vortex topological sector can 
be defined and the super-selection rule of Eq.~(\ref{eq:hilbert}) is
not realized. 

On the other hand, assuming that the 
scaling of the string tension $\sigma$ with the lattice spacing $a$, known 
analytically for $\beta_A=0$: 
\begin{eqnarray}
\beta_F &=& \frac{4}{a^2\, g^2}
\nonumber\\
\sigma &=& \frac{3}{8}\, g^2\,,
\label{eq:a_scaling_2}
\end{eqnarray} 
will hold up to a different prefactor along any line
$f(\beta_F)\propto \beta_F$, we get:
\begin{equation}
V = (a\,L)^2 = \frac{3}{2} \frac{L^2}{\sigma\,\beta_F}\,.
\end{equation} 
Keeping now the
volume $V$ fixed as the continuum limit is approached, the values of
the coupling at which one needs to simulate for fixed $L$ will scale
as $\beta_F \sim {L^2}$, i.e. much higher than the pseudo-critical coupling
$\beta^c_F \sim \log{L}$. The theory will thus be in a pseudo-ordered phase 
with $\langle {{z}} \rangle=1$: 
on a {\it finite} Euclidean $d=2$ torus the Wilson action can admit well 
defined $\mathbb{Z}_2$ topological sectors.
\footnote{Some interpretation issues of course arise in this case. E.g., 
speaking of zero temperature for a compactified, periodic
time is at best misleading. Of course, one could also consider
the case $L^2_s \sim \beta_F$, but to fix the temperature independently one 
must resort to an anisotropic (Hamiltonian)
setup \cite{Burgio:2003in}. Transitions on finite toruses in the large $N$ limit
of the $d=2$ Yang-Mills theories have been the subject of intense 
investigations; see Refs.~\cite{Narayanan:2007dv,Hietanen:2012ma} and 
references therein.}

\subsection{$d=3$}

Increasing the dimensions to $d =3$ we expect interactions to arise among 
parallel planes, since vortices are now extended, one-dimensional objects. The 
simple picture we have found in $d=2$ won't probably work anymore and less 
trivial critical exponents might arise. Still, fluxes are inherently 
two-dimensional objects and most of the dynamics should thus take place on 
planes: 
many features of the $d=2$ case should therefore survive. To check this, we 
have performed 
sets of Monte-Carlo simulations along different lines in the 
$\beta_F - \beta_A$ plane.
Results are reported for $\beta_A = 0$, $\beta_F = 0.5$, $0.75$ and lattice 
sizes between $L=24$ and $L=80$; other parameters have been checked and give
a consistent picture.

In the $\beta_A = 0$ case approximately 20 to 50 
simulations at coupling steps $\delta \beta_F$, each with $10^6$ independent 
configurations, were performed for each volume $L^3$. 
The data have been re-weighted \cite{Ferrenberg:1988yz,Ferrenberg:1989ui} 
to determine the peak values $\beta^c_F(L)$ and $\chi_{{z}}(\beta^c_F(L))$; 
this was viable only up to $L \sim 64$. 
Indeed, as we shall see below, the $d=3$ case shows a similar
scaling behaviour as Eq.~(\ref{eq:ffs_2d_F}),
i.e. a logarithmic scaling of $\beta_F^c(L)$ to a critical coupling
$\beta_F^c = \infty$. This has a practical drawback: 
the absolute width of the transition, i.e. the overall interval 
$\Delta \beta_F$ one needs to simulate, varies very slowly, while
the step-width $\delta \beta_F$ one must scan in order to keep the density
of states computationally feasible decreases dramatically with $L$:
the computational cost becomes eventually unmanageable. 

\begin{figure}
\begin{center}
\includegraphics[width=0.85\textwidth]{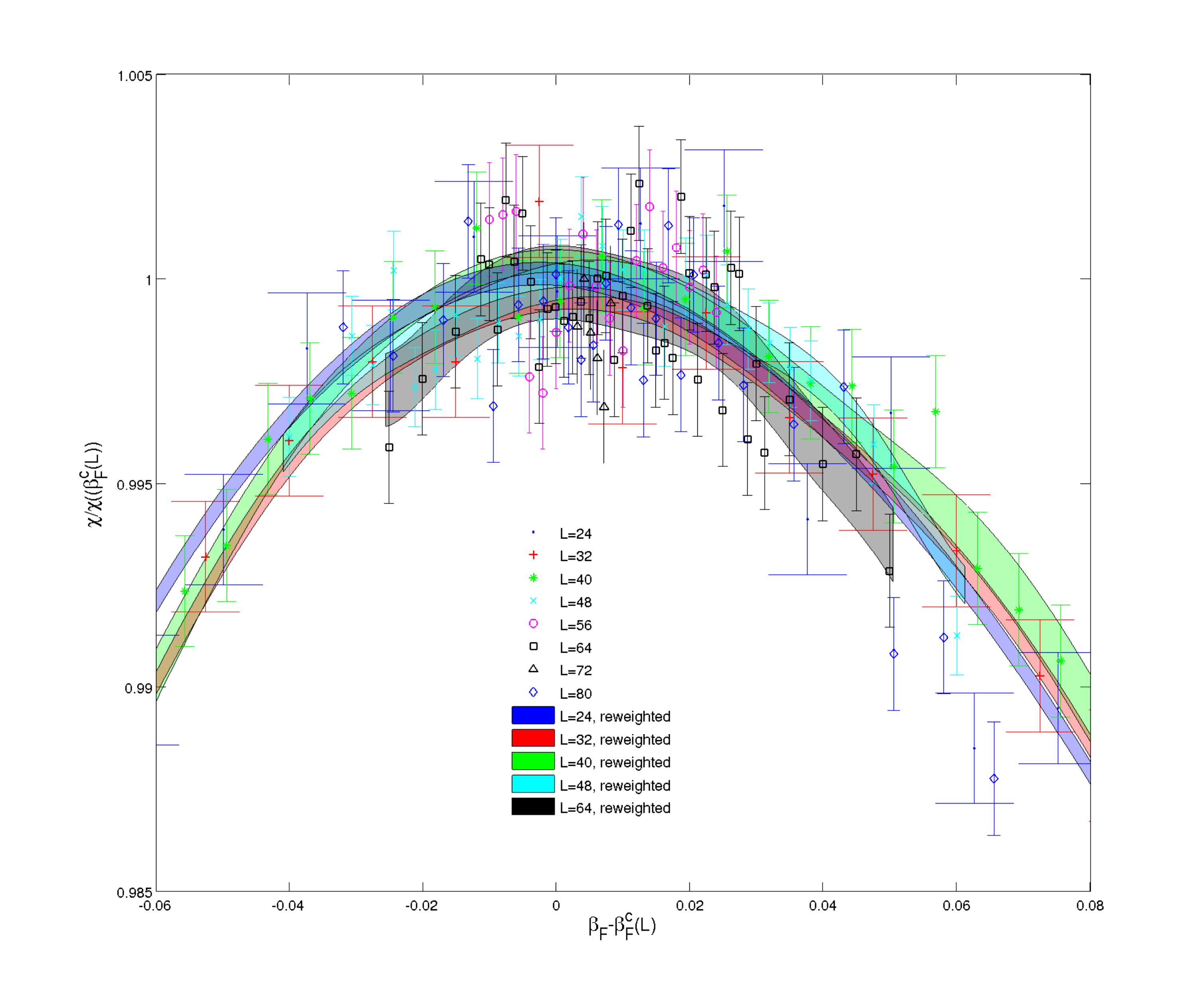}
\end{center}
\caption{Data for the susceptibility of the order parameter in 
$d=3$, 
including the re-weighted curves, rescaled 
with the FSS Ansatz in Eqs.~(\protect\ref{eq:fss_chi}, 
\protect\ref{eq:fss_beta_3d}).}
\label{fig_3}
\end{figure}

\begin{figure}
\begin{center}
\includegraphics[width=0.7\textwidth]{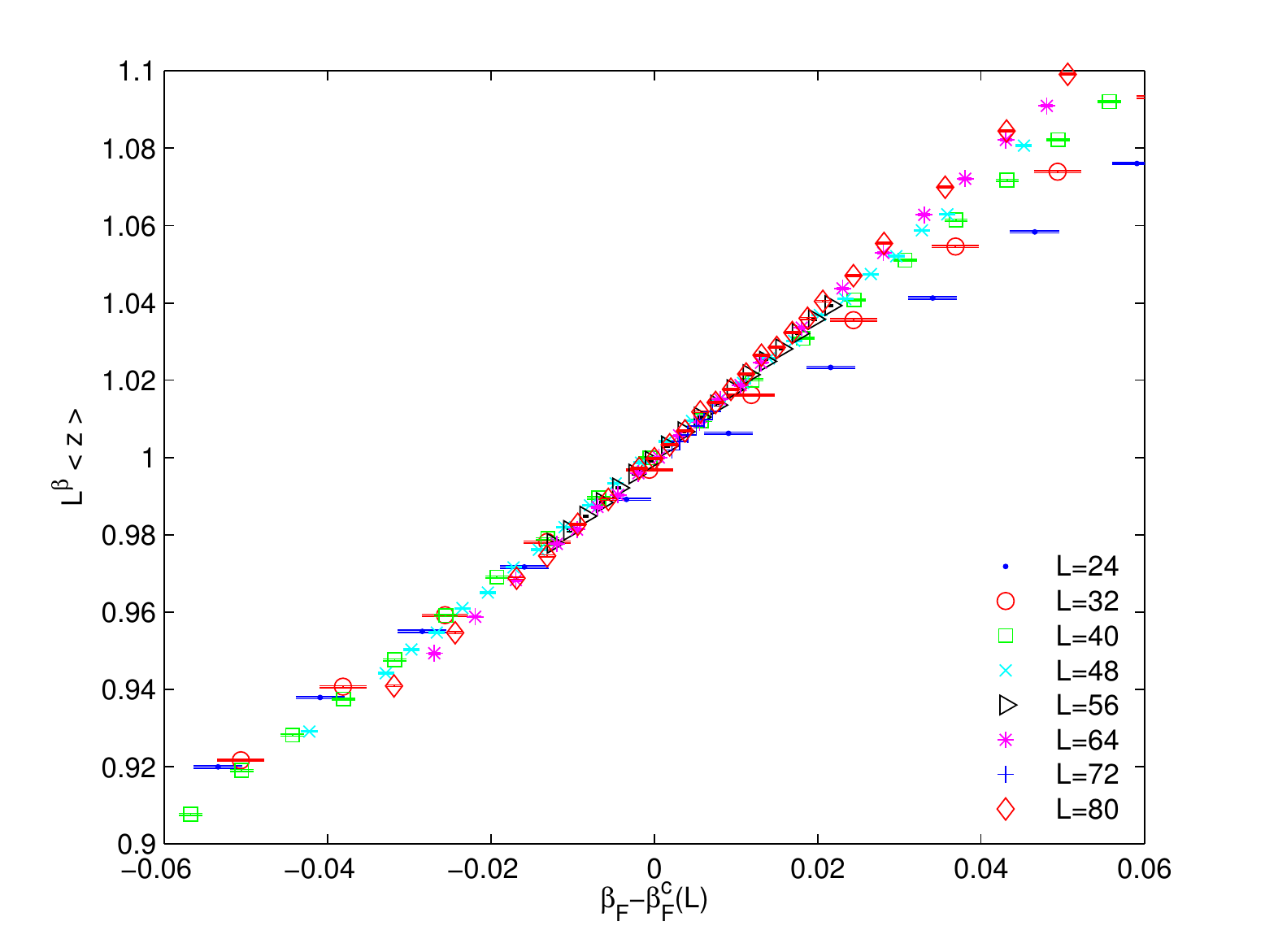}
\end{center}
\caption{FSS for the order parameter ${{z}}$ as a function of the rescaled 
coupling as in
Eq.~(\protect\ref{eq:fss_z}).}
\label{fig_33}
\end{figure}

Results for all volumes considered are resumed in Tab.~\ref{tab:2}, where 
the steps $\delta \beta_F$ are also listed, along with the value of the
center flux and, for sake of completeness, of the specific heat at the 
pseudo-critical point. To cross-check scaling results, 
similar simulation steps and statistics have also been used for the other 
volumes not included in the re-weighting.
\begin{table}
\begin{tabular}{|l|l|l|l|l|l|}
\hline
&$L=24$&$L=32$&$L=40$&$L=48$&$L=64$\tabularnewline
\hline
$\beta^c_F(L)$&$5.61(1)$&$5.930(5)$&$6.174(2)$&$6.383(5)$&
$6.700(5)$\tabularnewline
$\chi_{{z}}(\beta^c_F(L))$&$152.90(4)$&$282.83(8)$&$454.24(12)$&$666.6(2)$&
$1217.4(3)$\tabularnewline
${{z}}(\beta^c_F(L))$&$0.392(3) $&$0.3610(15)$&$0.3364(13)$&$0.3213(16)$&
$0.2908(16)$\tabularnewline
$C_s(\beta^c_F(L))$&$0.1078(3)$&$0.0957(1)$&$0.0877(2)$&$0.0817(3)$&
$0.0739(4)$\tabularnewline 
$\delta \beta_F$&$0.025$&$0.0125$&$0.00625$&$0.003125$&
$0.0015625$\tabularnewline
\hline
\end{tabular}
\caption{Position and height of the susceptibility peaks along $\beta_F$ in 
$d=3$. The third and fourth line give the values of the order parameter and 
of the specific heat at the pseudo-critical point $\beta^c_F(L)$; the last 
line gives the coupling steps for the simulations used in the re-weighting.}
\label{tab:2}
\end{table}
The data can be well fitted with the Ansatz:
\begin{eqnarray}
\chi_{{z}}(\beta^c_F(L)) &\sim &A\, L^{2-\eta} \log^{-2 r}{L}\left(1
+{\cal{O}}(L^{-1})\right)\label{eq:fss_chi}\\
{{z}}(\beta^c_F(L)) &\sim &M \, L^{-\beta} \left(1+{\cal{O}}({L}^{-1}) 
\right)\label{eq:fss_z}\\
\beta^c_F(L) &\sim& C\,\log{L^2}+D\,\log\log{L^2}
+{\cal{O}}(1)\label{eq:fss_beta_3d}\,,
\end{eqnarray}
For $\chi_{{z}}$ we get $A= 0.21(1)$, $r=-0.134(10)$, %$B=-0.4101$ 
$\eta = 0.0001(100)$ and $\chi^2$/d.o.f.$= 5.7$; constraining $\eta=0$
gives again $A = 0.21(1)$, $r = -0.134(10)$ with 
%$B = -0.4106$ 
$\chi^2$/d.o.f.$ = 2.8$.
For $\beta_F^c$ we get 
$C= 0.61(3)$ and $D=-0.42(5)$ with %$E=2.5034$ 
$\chi^2$/d.o.f.$= 0.7$; on the other hand, constraining $D=0$ 
we get $C=0.56(3)$, 
%$E=2.0738$ 
$\chi^2$/d.o.f.$= 0.6$. Finally, for the order parameter, we get $M= 1.26(5)$ 
and 
$\beta=0.35(1)$ with $\chi^2$/d.o.f.$= 1.4$.
Overall, the biggest source of systematic error
is given by the parameterization of the sub-leading corrections:
leaving them out or parameterizing them differently leads to 
changes of up to $10\%$ for some 
of the critical exponents, not included in our error estimates; 
obviously, more data at higher volumes 
are needed to pin the numbers down.\footnote{Another 
possible issue could be the non-ergodicity of our 
set-up in the ordered phase. Indeed, a ``good'' algorithm would need to change 
boundary conditions to enable tunneling among different topological
sectors around the transition, just like a cluster algorithm in an Ising model
allows tunneling among different orientations of the spins in the 
spontaneously magnetized phase. See e.g. Ref.~\cite{vonSmekal:2012vx} for
possible solutions to the problem. Implementing such algorithm is obviously
beyond the scope of this paper.}
The data for the susceptibility $\chi_{{z}}$, rescaled
by Eqs.~(\protect\ref{eq:fss_chi},
\protect\ref{eq:fss_beta_3d}), are plotted in Fig.~\ref{fig_3},
showing very good agreement also
for the volumes which have not been included in the re-weighting analysis. 
In Fig.~\ref{fig_33} we show the scaling of the order parameter ${{z}}$ 
according to Eqs.~(\protect\ref{eq:fss_z}, \protect\ref{eq:fss_beta_3d}); 
the agreement for $L \gtrsim 40$ is again very good. 
As for the specific heat, a fit of the data in Tab.~\ref{tab:2} with a 
logarithmic Ansatz $C_s \sim \log^{-\alpha}{L}$ gives $\alpha = 1.4(1)$ with
a $\chi^2$/d.o.f.$= 0.5$. The signal-to-noise ratio for the MC is however 
not so good in this case, reflecting itself in the quality of the 
re-weighted data: more statistics would be definitely needed; anyway, 
checking any (hyper-) scaling relation is beyond our goals.

The above result is quite surprising. Indeed, in contrast to $d=2$, one could 
have expected the $\mathbb{Z}_2$ monopole to control the 
open center vortices, since the density of the latter is proportional
to that of the former. However, although monopoles per unit volume steadily 
decrease beyond the cross-over, open vortices ``connecting'' them still cause 
a critical 
behaviour cumulating to $g^2 \to 0$.\footnote{A somewhat cryptic comment
regarding a possible critical behaviour, going as far as taking the 
$XY$-model as a paradigm, can be found in Ref.~\cite{Mack:1979gb}.}
A possible explanation could be that 
their length increases more than linearly with the lattice size; multiple
bendings in orthogonal directions would be enough to randomize the fluxes. 
A direct 
investigation of any geometrical properties of open vortices is however beyond 
the scope of this paper, since Eq.~(\ref{eq:twist}) is non local and gauge 
invariant and does not allow to isolate the topological defects on the planes.

\begin{figure}
\begin{center}
\includegraphics[width=0.7\textwidth]{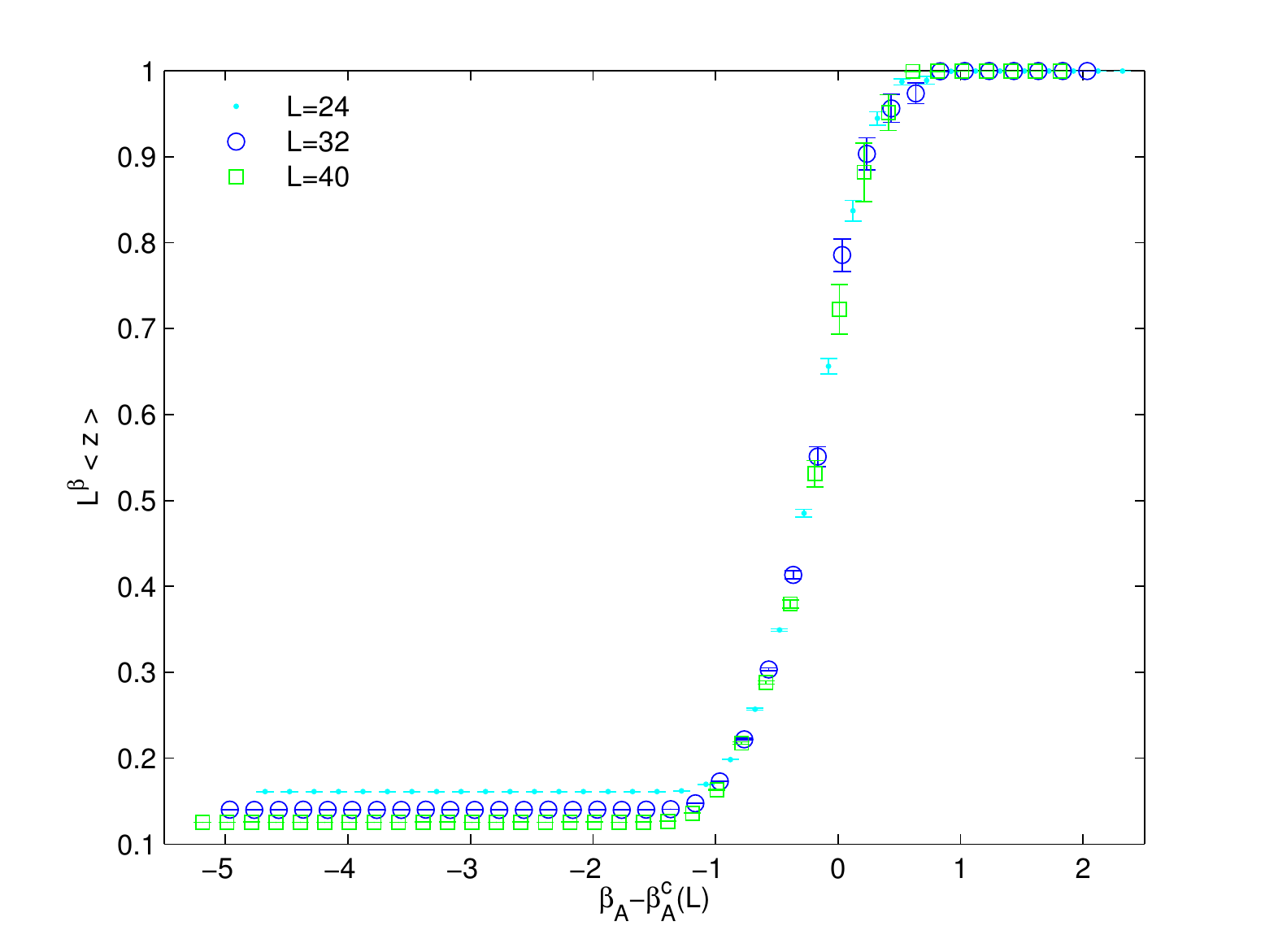}
\end{center}
\caption{FSS for the order parameter ${{z}}$ along the 
$\beta_F = 0.5$ line with the Ansatz in Eqs.~(\protect\ref{eq:fss_3d_A_1},
\protect\ref{eq:fss_z_3}).}
\label{fig:z_3d_A}
\end{figure}

\begin{figure}
\begin{center}
\includegraphics[width=0.7\textwidth]{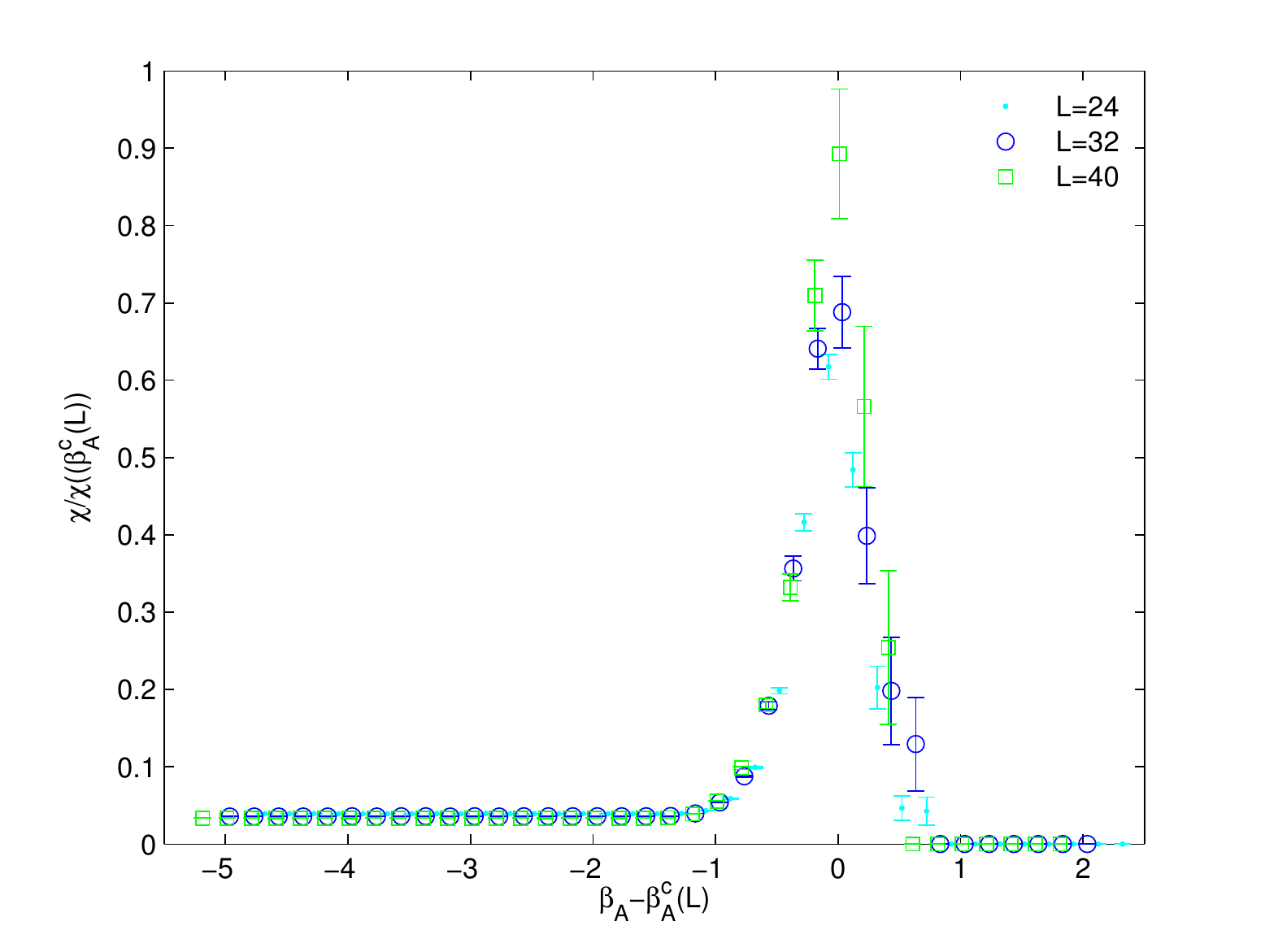}
\end{center}
\caption{FSS for the susceptibility $\chi_{{z}}$ along the 
$\beta_F = 0.5$ line with the Ansatz Eqs.~(\protect\ref{eq:fss_3d_A_1},
\protect\ref{eq:fss_3d_A}).}
\label{fig_4}
\end{figure}

\begin{figure}
\begin{center}
\includegraphics[width=0.7\textwidth]{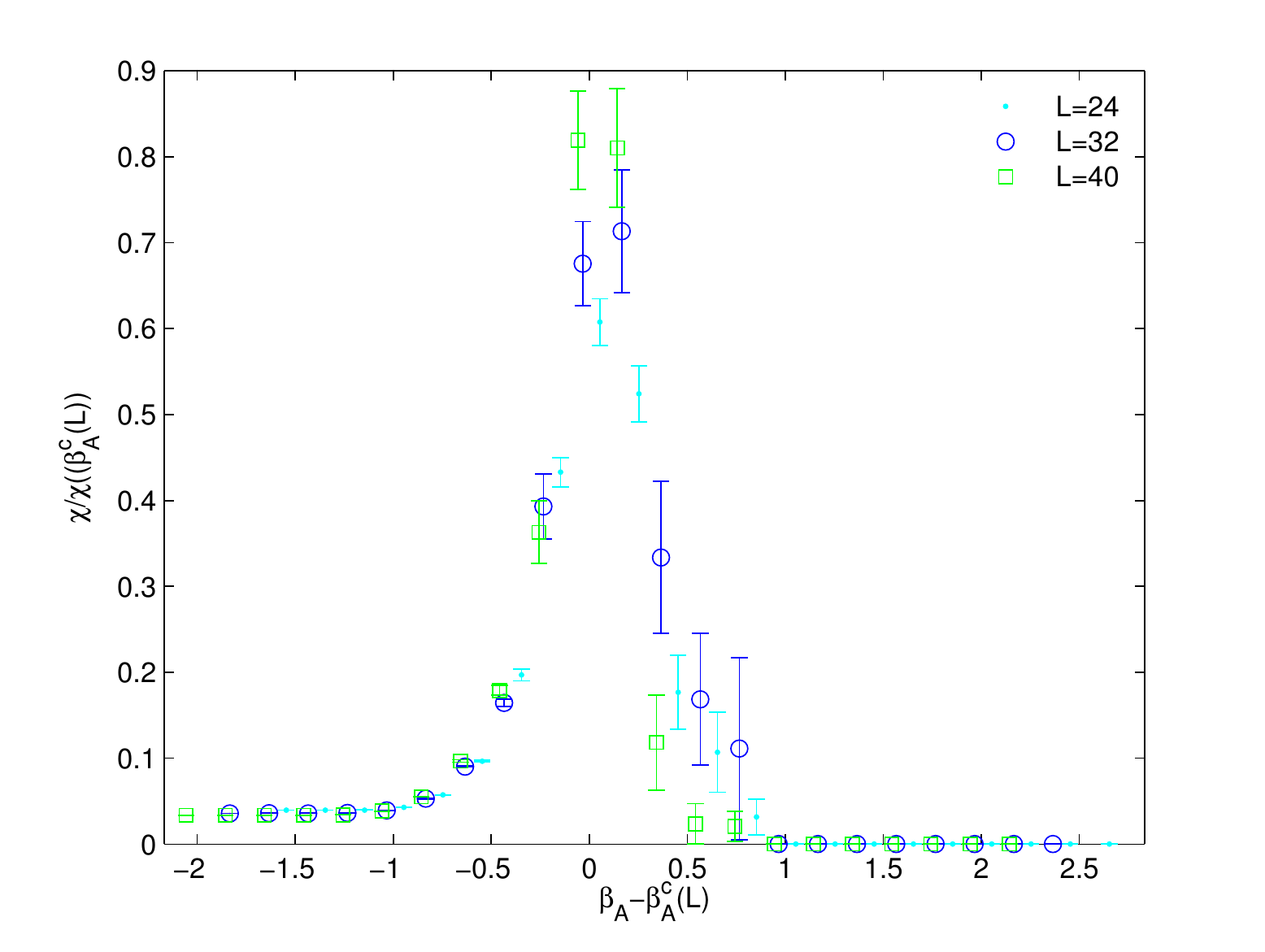}
\end{center}
\caption{FSS for the susceptibility $\chi_{{z}}$ along the 
$\beta_F = 0.75$ line with the Ansatz Eqs.~(\protect\ref{eq:fss_3d_A_1},
\protect\ref{eq:fss_3d_A}).}
\label{fig_4a}
\end{figure}

Going now to the $\beta_A \neq 0$ case, since the $\mathbb{Z}_2$ 
monopoles undergo a cross-over also in the low 
$\beta_F$ region of the phase diagram of Fig.~\ref{fig:plane3}, 
one would expect the center flux to behave as in the $\beta_F$ case:
one should find along $\beta_A$ a similar scaling as in 
Eqs.~(\ref{eq:fss_chi}, \ref{eq:fss_beta_3d}).
Also, the transition lines should not be effected by the bulk transition 
associated with the unphysical $\mathbb{Z}_2$ gauge degrees of freedom. However,
such strong transition unavoidably makes any simulation near it quite noisy; 
on top of that the biased Metropolis algorithm, with e.g. 3 micro-canonical 
steps, gets 
inefficient as $\beta_F$ gets small and $\beta_A$ large, reaching 
for ${{z}}$ and $\chi_{{z}}$, around the peaks of the latter, integrated 
autocorrelation 
times of the order $10^4 - 10^5$ for $24 \leq L \leq 40$.
Passable data were therefore only accessible for three volumes, while 
gathering 
enough statistics to re-weight the susceptibility was out 
of the question. We have thus limited ourself to a consistency check 
near the bulk transition with a scaling Ansatz similar to 
Eqs.~(\ref{eq:fss_chi}-\ref{eq:fss_beta_3d}):
\begin{eqnarray}
\chi_{{z}}(\beta^c_A(L)) &\sim &A\, L^{2} \log^{-2 r}{L}+{\cal{O}}(L)
\label{eq:fss_3d_A_1}\\
{{z}}(\beta^c_A(L)) &\sim &M \, L^{-\beta} \left(1+{\cal{O}}({L}^{-1}) \right)
\label{eq:fss_z_3}\\
\beta^c_A(L) &\sim& C\,\log{L^2}+{\cal{O}}(1)\,;
\label{eq:fss_3d_A}
\end{eqnarray}
no fit has been attempted. The scaling of the pseudo critical point 
and of the order parameter are consistent with the $d=2$, $\beta_A = \infty$ 
case, $C=1/2$ and $\beta=0$, as can be seen from Fig.~\ref{fig:z_3d_A}.
On the other hand, the peaks of $\chi_{{z}}$ are quite noisy and even a 
consistency check for the logarithmic exponent is hopeless.
In Fig.~\ref{fig_4}, \ref{fig_4a} we show the results for the simulations along 
$\beta_F=0.5$ and $\beta_F=0.75$, lying respectively left and right of the 
bulk transition, rescaled by Eq.~(\ref{eq:fss_3d_A_1}, \ref{eq:fss_3d_A}) with 
a ``guessed'' value for $r=-1/2$; of course, further work would be needed to 
determine the critical exponents reliably. 

We have also checked via Monte-Carlo simulations that
all of the above results generalize to $T>0$ by simply substituting
$L^2 \to L_s \cdot L_t$ in all the scaling relations for temporal fluxes,
while the behaviour of all spacial fluxes remains unchanged.
Again, as in the $d=2$ case, this implies that, along any line in 
the $\beta_F - \beta_A$ plane, when taking the thermodynamic limit before the 
weak coupling limit, i.e. sending the volume to infinity, the $d=3$ 
theory remains stuck in the disordered phase $\langle {{z}} \rangle = 0$;
again, Eq.~(\ref{eq:hilbert}) is not realized. 

What about the fixed volume limit? Taking as a blueprint for the continuum 
limit along any direction the scaling of 
the string tension along $\beta_F$ \cite{Teper:1998te}:
\begin{eqnarray}
\beta_F &=& \frac{4}{a\, g^2}
\nonumber\\
a \, \sqrt{\sigma} &=& \frac{c_0}{\beta_F } + \frac{c_1}{\beta^2_F } 
+{\cal{O}}(\frac{1}{\beta^3_F })\,.
\label{eq:a_scaling_3}
\end{eqnarray}
we get immediately:
\begin{equation}
V = (a\,L)^3 \propto \frac{L^3}{\beta_F^3 \sqrt{\sigma^3}}\,.
\end{equation}
Keeping again $V$ fixed as the continuum limit is 
approached, the values of the coupling corresponding to a given $L$ will now 
scale as $\beta_F \sim {L}$; again, as in $d=2$, they will always be much 
higher than the pseudo-critical coupling $\beta^C_F \sim \log{L}$ and
the Wilson action could admit well defined $\mathbb{Z}_2$
topological sectors on a {\it finite} $d=3$ torus. 

\subsection{$d=4$}

\begin{figure}
\begin{center}
\includegraphics[width=0.9\textwidth]{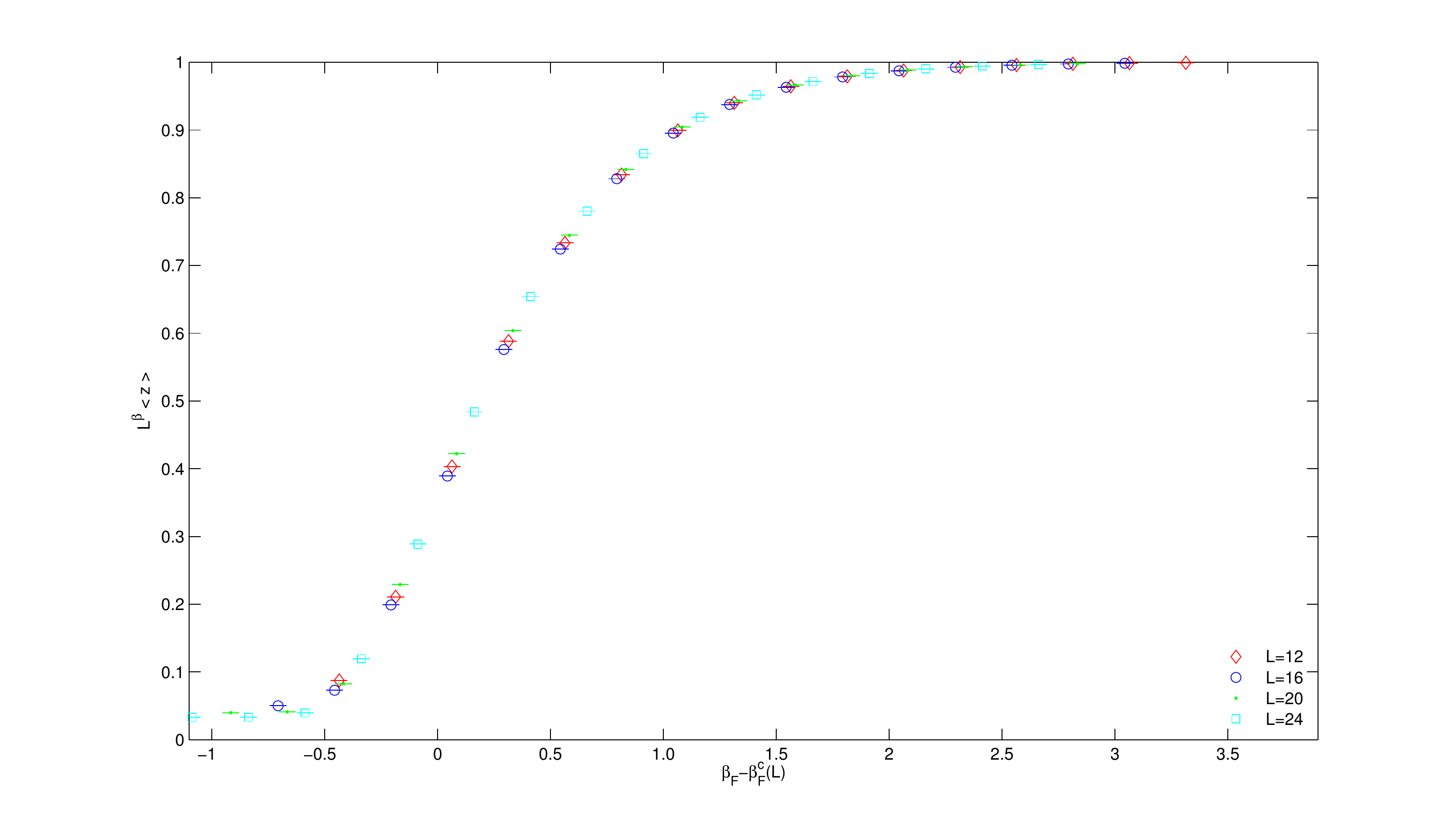}
\end{center}
\caption{FSS as in Eqs.~(\protect\ref{eq:fss_z_4}, 
\protect\ref{eq:fss_bt_4}) for ${{z}}$ in $d=4$ along the $\beta_F$ axis.}
\label{fig_5a}
\end{figure}

\begin{figure}
\begin{center}
\includegraphics[width=0.9\textwidth]{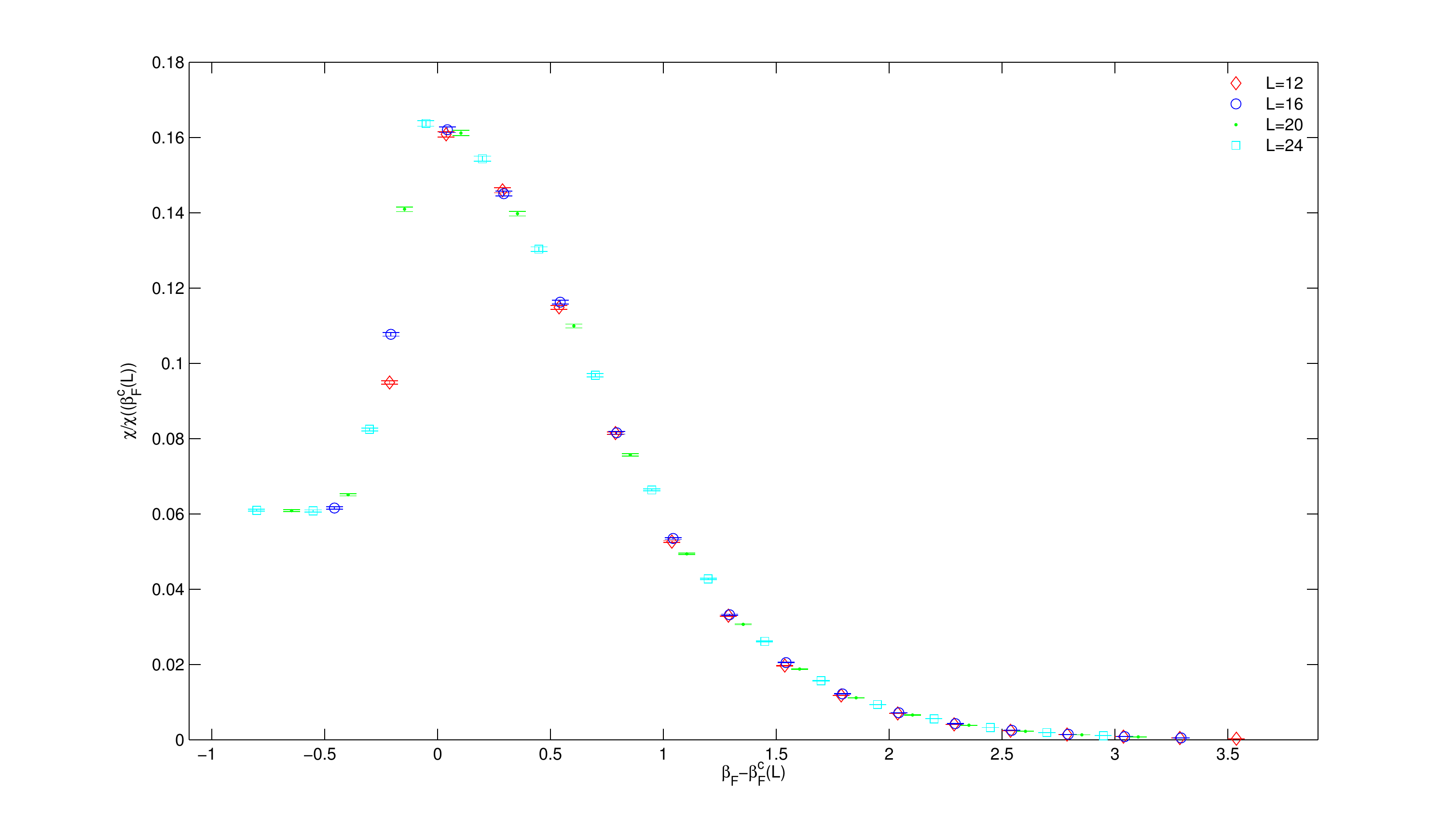}
\end{center}
\caption{FSS as in Eqs.~(\protect\ref{eq:fss_chi_4}, 
\protect\ref{eq:fss_bt_4}) for $\chi_{{z}}$ in $d=4$ along the $\beta_F$ axis.}
\label{fig_5}
\end{figure}

\begin{figure}
\begin{center}
\includegraphics[width=0.7\textwidth]{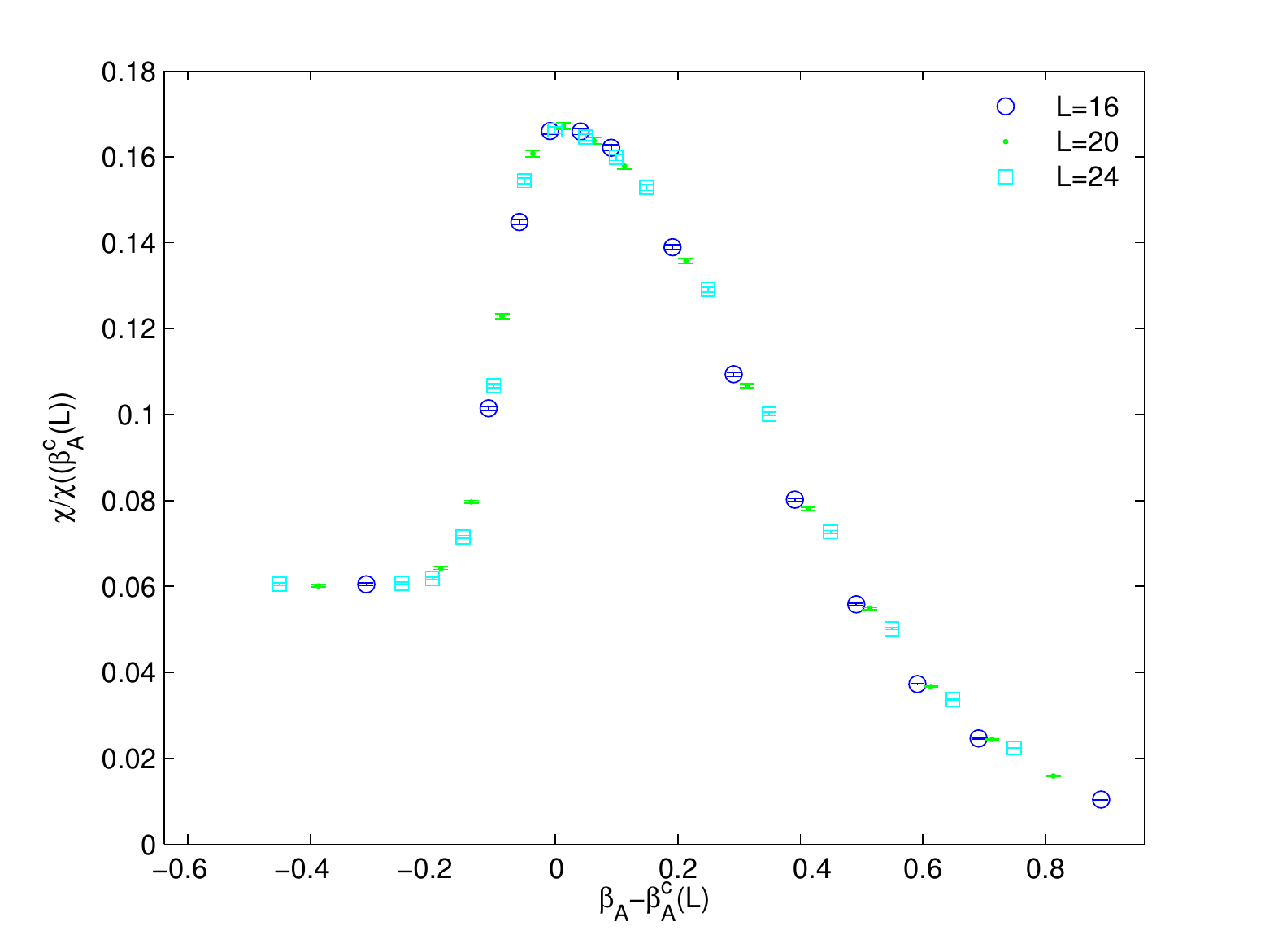}
\end{center}
\caption{FSS as in Eqs.~(\protect\ref{eq:fss_chi_4}, 
\protect\ref{eq:fss_bt_4}) for $\chi_{{z}}$ in $d=4$ along the $\beta_F = 1.3$ axis.}
\label{fig_6}
\end{figure}

The positions of the peaks of $\chi_{{z}}$, 
as obtained in the simulations along the $\beta_A = 0$, $\beta_F = 1.0$, 
$\beta_F = 1.2$ and $\beta_F = 1.3$ lines, all within phase I of 
Fig.~\ref{fig:plane4}, are shown in Tabs.~\ref{tab:3}, 
\ref{tab:4}.
\begin{table}
\begin{tabular}{|l|c|c|c|c|c|}
\hline
&$L=12$&$L=16$&$L=20$&$L=24$\tabularnewline
\hline
$\beta^c_F(L)$&$3.15(5)$&$3.40(5)$&$3.60(5)$&$3.75(5)$\tabularnewline
$\chi_{{z}}(\beta^c_F(L))$&$24(1)$&$42(1)$&$66(1)$&$95(1)$\tabularnewline
\hline
\end{tabular}
\caption{Position and height of the susceptibility peaks along $\beta_F$ in 
$d=4$.}
\label{tab:3}
\end{table}
\begin{table}
\begin{tabular}{|l|c|c|c|c|c|}
\hline
&$L=16$&$L=20$&$L=24$\tabularnewline
\hline
$\beta_F = 1.0 $&$1.845(25)$&$1.923(25)$&$1.987(25)$\tabularnewline
$\beta_F = 1.2 $&$1.695(25)$&$1.773(25)$&$1.837(25)$\tabularnewline
$\beta_F = 1.3 $&$1.62(1)$&$1.70(10)$&$1.76(1)$\tabularnewline
\hline
\end{tabular}
\caption{Position of the susceptibility peaks along $\beta_A$ in 
$d=4$ for $\beta_F = 1.0$, 1.2 and 1.3; the heights are all compatible
with the results in Tab.~\protect\ref{tab:3}}
\label{tab:4}
\end{table}
We have again limited ourselves to a consistency check with a scaling Ansatz of 
the form:
\begin{eqnarray}
\chi_{{z}}(\beta^c_F(L)) &\sim &A\, L^{2} +{\cal{O}}(L)\label{eq:fss_chi_4}
\\
{{z}}(\beta^c_F(L)) &\sim &M \, L^{-\beta} \left(1+{\cal{O}}({L}^{-1}) \right)
\label{eq:fss_z_4}\\
\beta^c_F(L) &\sim& C\,\log{L^2}+{\cal{O}}(1)\;.
\label{eq:fss_bt_4}
\end{eqnarray}
Results are shown in Figs.~\ref{fig_5a}, \ref{fig_5} for the order
parameter and its susceptibility along the $\beta_F$ axis; up to the values 
of $C$ the behaviour along 
the lines parallel to the $\beta_A$ axis is basically the same, see e.g. 
Fig.~\ref{fig_6}. From the data in Tab.~\ref{tab:3} we can estimate 
$C=0.46(3)$, compatible with $1/2$, while for those in Tab.~\ref{tab:4} we 
get $C=0.18(3)$; in all cases $\beta$ is compatible with 0.

Direct simulations at $T>0$ give again the same scaling with 
$L^2 \to L_s \cdot L_t$ for the temporal fluxes, while spacial fluxes
remain unchanged. As in the $d=2$ and $d=3$ 
cases, in the 
thermodynamic limit the theory remains therefore
stuck in the disordered phase. Moreover, starting from the 2-loop 
beta-function, the running of the physical scale with 
$\alpha_{\mathrm{lat}} = g^2/(4\,\pi)$ is given by:
\begin{equation}
\log{\left(a^2 \, \sigma\right)} = 
-\frac{4\,\pi}{\beta_0}\,\alpha_{\mathrm{lat}}^{-1} + 
\frac{2\,\beta_1}{\beta_0^2}\,\log{\left(\frac{4\,\pi}{\beta_0}\,
\alpha_{\mathrm{lat}}^{-1}\right)} + c + {\cal{O}}(\alpha_{\mathrm{lat}})\,,
\label{eq:a_scaling_4}
\end{equation}
where $c = \log{\frac{\sigma}{\Lambda^2_{\mathrm{lat}}}}$ and $\beta_0 = 
\frac{22}{3}$, $\beta_1 = \frac{68}{3}$, i.e. from Eq.~(\ref{eq:mixed}):
\begin{equation}
V = (a\,L)^4 \propto {L^4}\,\left(\frac{4\,\pi^2}{\beta_0}\,
\beta_F\right)^{\frac{4\,\beta_1}{\beta_0^2}}\,\mathrm{e}^{-\frac{8\,\pi^2}{\beta_0}\,\beta_F}\,.
\end{equation}
When trying to keep the volume $V$ fixed as $\beta_F \to 0$, up to 
$\log\log$ corrections the coupling should scale as: 
\begin{equation}
\beta_F \sim \frac{\beta_0}{4\,\pi^2}\,
\log{L^2}\,;
\label{scale_4d}
\end{equation} 
the coefficient $C$ in 
Eq.~(\ref{eq:fss_bt_4}) is however larger than that coming from the 
beta-function and the simulation parameters will lie in the disordered phase: 
topological sectors will always be 
ill-defined also on a finite torus $T^4$. The same holds along the lines 
parallel to the $\beta_A$ axis (see Tab.~\ref{tab:4}); in this case, from 
Eq.~(\ref{eq:mixed}), the coefficient in Eq.~(\ref{scale_4d}) coming from the 
beta-function Eq.~(\ref{eq:a_scaling_4}) is
${3\,\beta_0}/({32\,\pi^2})$, again smaller than the corresponding value 
of $C$.

As discussed in Sec.~\ref{sec:alg}, we have excluded phase II (see 
Fig.~\ref{fig:plane4}) from the simulations. As mentioned above, vortex topology
is however well understood in this case: the results of 
Refs.~\cite{Barresi:2003jq,Barresi:2004qa,Barresi:2006gq,Burgio:2006dc,%
Burgio:2006xj} show that, in contrast to phase I in $d=4$ and to 
the $d=2$ and 3 cases, the theory possesses well defined $\mathbb{Z}_2$ 
topological sectors in the continuum limit. 

\section{Conclusions}
\label{sec:con}

We have studied a topological order parameter,
the center flux $z$ defined in Eqs.~(\ref{eq:op}, \ref{eq:op2_al}), 
for the $SU(2)$ mixed action in $2 \leq d \leq 4$.
Its ordered phase, $\langle z \rangle = 1$, corresponds to well defined
$\pi_1(SO(3))=\mathbb{Z}_2$ topological sectors, i.e. to a vacuum 
satisfying the super-selection rule of Eqs.~(\ref{eq:hilbert}, \ref{eq:vac}),
while for $\langle z \rangle = 0$ the vacuum state is disordered and no 
center topology can be defined. This reminds of a quantum phase transition; 
however, one does not switch between vacua by
tuning a physical parameter. Rather, the choice of dimensions and the symmetry
of the discretized action control in which phase the theory will be in 
the continuum limit. 

More specifically, discretized actions transforming in the fundamental 
representation possess a disordered vacuum, with $z$ showing an essential 
scaling to the critical coupling $\beta_c = \infty$. 
The critical exponent for the correlation length $\xi$ is $\nu=1$, i.e. 
$\beta_c(\xi) \propto \log{\xi}$; explicit $\log \log{\xi}$ corrections 
to scaling can be shown to exist for some choice of parameters. 
The susceptibility of the center flux scales as $\chi_{{z}}(\xi)  \propto \xi^2$ 
in $d=2$ and 
$d=4$, while the order parameter itself scales trivially in these cases.
On the other hand in $d=3$, at least along the $\beta_F$ 
axis, the center flux has a non trivial critical exponent, 
${{z}}(\xi) \propto \xi^{-\beta}$, with $\beta=0.35(1)$, while
a logarithmic correction can be explicitly determined 
for the scaling of its susceptibility, 
$\chi_{{z}}(\xi) \sim \xi^2\,\log^{-2r}{\xi}$, with $r=-0.134(10)$; 
similar corrections might also be present along other lines in the 
$\beta_F - \beta_A$ diagram, but more statistics would be needed to reach a 
conclusive result. A tentative critical exponent for the specific heat,
$C_s(\xi) \sim \log^{-\alpha}{\xi}$, gives $\alpha = 1.4(1)$, but with
still high systematic errors. We have made no attempt to investigate any
(hyper-) scaling relations among such exponents; this would probably require a
full analysis of the Lee-Yang zeros \cite{Kenna:1996bs}.
Such behaviour persists in all dimensions at $T>0$.

Vice versa, the topological classification of Eq.~(\ref{eq:homotopy})
and thus the super-selection rule of Eqs.~(\ref{eq:hilbert}, \ref{eq:vac})
can be realized by the vacuum state of lattice actions transforming in the 
adjoint representation; phase II in $d=4$ (see Fig.~\ref{fig:plane4}) is such 
an example \cite{Burgio:2006dc,Burgio:2006xj,Barresi:2006gq}. 
Large scale simulations with the adjoint action are hampered by strong 
finite-volume effects \cite{Halliday:1981te,Halliday:1981tm,deForcrand:2002vs}. 
Therefore, although the techniques used in 
Refs~\cite{Burgio:2006dc,Burgio:2006xj,Barresi:2006gq} 
to tame them could also work in $d=3$, a more viable 
alternative, applicable also in $d=2$, would be to resort to positive 
plaquette models \cite{Mack:1981gy,Bornyakov:1991gq,Fingberg:1994ut},
where topological sectors are always well defined since the 
operator given in Eq.~(\ref{eq:twist}) takes ``by construction'' the 
values dictated by the assigned boundary conditions. Indeed, a
one-to-one mapping between configurations in such lattice discretization 
and those of the adjoint Wilson action with well defined vortex sectors was
conjectured in Ref.~\cite{deForcrand:2002vs} and explicitly constructed in 
Ref.~\cite{Barresi:2006gq}. Finally, an ordered vacuum could also be realized 
for a finite torus in $d=2,\,3$; here one could exploit the power-law scaling 
of the physical mass with the 
coupling to define topological sectors when $L\to \infty$ and 
$a \to 0$ with the volume $V= (a\,L)^d$ kept fixed.

The above findings do not contradict universality, since non perturbatively 
the equivalence between fundamental and adjoint actions 
can only hold as long as no lattice artifacts are present 
\cite{Mack:1978rq,Mack:1979gb,Halliday:1981te,%
Halliday:1981tm,Coleman:1982cx,deForcrand:2002vs,Barresi:2006gq}, while 
as we have seen for some discretizations the density of 
$\mathbb{Z}_2$ monopoles can not vanish at any finite coupling 
\cite{Mack:1979gb}.
Does however such result have any physical consequences? The vacua 
of the two different phases can be essentially characterized by the type of 
$\mathbb{Z}_2$ vortices they can carry:

\noindent i) The ordered phase allows topological center vortices ``\`a la 
't Hooft'' \cite{'tHooft:1977hy,'tHooft:1979uj}: a confinement mechanism
based on the super-selection rule of Eqs.~(\ref{eq:hilbert}, \ref{eq:vac})
can be realized; at finite temperature the change in the vortex free energy 
as measured via Eq.~(\ref{eq:free_en}) is thus a valid test to establish how 
the symmetry is broken in the transition to the deconfined phase 
\cite{Burgio:2006dc,Burgio:2006xj,Barresi:2006gq}. No fundamental
fields are allowed in this case 
\cite{'tHooft:1977hy,'tHooft:1979uj,Cohen:2014swa}; however, adjoint fermions 
can be easily incorporated in such scenario. It might therefore 
be interesting to investigate the vacuum properties of the $SU(2)$ gauge 
theory coupled to adjoint fermions, a popular candidate 
for infrared conformality \cite{Lucini:2013wsa}. Numerical tests 
with the adjoint Wilson action or positive plaquette model should be viable.

\noindent ii) The disordered phase is dominated by (one huge, percolating?) 
open vortices, reminding of the Nielsen-Olesen ``spaghetti vacuum'' 
\cite{Nielsen:1973cs}.
Such open vortices are not topological according to Eq.~(\ref{eq:homotopy}):
Eqs.~(\ref{eq:hilbert}, \ref{eq:vac}) cannot be applied. One might 
conjecture some relationship with P-vortices 
\cite{DelDebbio:1996mh,Langfeld:1997jx,Greensite:2011zz}, although it is 
still unclear how to test such hypothesis, since the center flux $z$ is gauge 
invariant and constructed out of pure $SO(3)$ variables while P-vortices are 
gauge dependent and built out of the $\mathbb{Z}_2$ gauge degrees of freedom. 
Moreover, such open vortices persist at any temperature, not disappearing
above $T_C$.
This disordered vacuum is detached from the boundary conditions 
chosen and is therefore compatible with the presence 
of fundamental matter fields.
Of course, Eq.~(\ref{eq:free_en}) is ill defined in this case; whether
any vortex related order parameter for the confinement-deconfinement phase 
transition could be defined remains an open question.

\section*{Aknowledgements}
We are indebted to F. Bursa for precious correspondence on the 
$d=2$ Yang-Mills theory. 
We want to thank B. Lucini, P. de Forcrand and 
D. Campagnari for critical reading of the manuscript;
we also wish to thank R. Kenna for interesting discussions on Lee-Yang
and Fisher zeroes.
A special thanks goes to E. Rinaldi for correspondence on the
implementation of the biased Metropolis algorithm.
This work was supported by the Bundesministerium f\"ur Bildung und Forschung 
under the contract BMBF-05P12VTFTF.

\bibliographystyle{apsrev4-1}
\bibliography{references}

\end{document}